\documentclass[12pt]{article}

\usepackage{amsmath,amssymb,amsthm,amscd,a4wide, amscd}
\usepackage{fancybox, ascmac}
\usepackage[hidelinks]{hyperref}

\usepackage{amsfonts,latexsym,makeidx}
\usepackage{amsxtra}
\usepackage{graphicx}
\usepackage{graphicx,color}
\usepackage{ulem} 
\usepackage{bm}
\usepackage{subcaption}

\makeatletter
\DeclareFontFamily{U}{tipa}{}
\DeclareFontShape{U}{tipa}{m}{n}{<->tipa10}{}
\newcommand{\arc@char}{{\usefont{U}{tipa}{m}{n}\symbol{62}}}%

\newcommand{\arc}[1]{\mathpalette\arc@arc{#1}}

\newcommand{\arc@arc}[2]{%
  \sbox0{$\m@th#1#2$}%
  \vbox{
    \hbox{\resizebox{\wd0}{\height}{\arc@char}}
    \nointerlineskip
    \box0
  }%
}
\makeatother

\begin{document}

\newcommand{\End}{{\rm{End}\ts}}
\newcommand{\Hom}{{\rm{Hom}}}
\newcommand{\Mat}{{\rm{Mat}}}
\newcommand{\ad}{{\rm{ad}\ts}}
\newcommand{\ch}{{\rm{ch}\ts}}
\newcommand{\chara}{{\rm{char}\ts}} 
\newcommand{\diag}{ {\rm diag}}
\newcommand{\pr}{^{\tss\prime}}
\newcommand{\non}{\nonumber}
\newcommand{\wt}{\widetilde}
\newcommand{\wh}{\widehat}
\newcommand{\ot}{\otimes}
\newcommand{\ls}{\ts\lambda\ts}
\newcommand{\La}{\Lambda}
\newcommand{\De}{\Delta}
\newcommand{\Ga}{\Gamma}
\newcommand{\vk}{\varkappa}
\newcommand{\vt}{\vartheta}
\newcommand{\si}{\sigma}
\newcommand{\vp}{\varphi}
\newcommand{\ze}{\zeta}
\newcommand{\om}{\omega}
\newcommand{\su}{s^{}}
\newcommand{\hra}{\hookrightarrow}
\newcommand{\ve}{\varepsilon}
\newcommand{\ts}{\,}
\newcommand{\vac}{\mathbf{1}}
\newcommand{\di}{\partial}
\newcommand{\qin}{q^{-1}}
\newcommand{\tss}{\hspace{1pt}}
\newcommand{\Sr}{ {\rm S}}
\newcommand{\U}{ {\rm U}}
\newcommand{\BL}{ {\overline L}}
\newcommand{\BE}{ {\overline E}}
\newcommand{\BP}{ {\overline P}}
\newcommand{\AAb}{\mathbb{A}\tss}
\newcommand{\CC}{\mathbb{C}\tss}
\newcommand{\KK}{\mathbb{K}\tss}
\newcommand{\QQ}{\mathbb{Q}\tss}
\newcommand{\SSb}{\mathbb{S}\tss}
\newcommand{\ZZ}{\mathbb{Z}\tss}
\newcommand{\X}{ {\rm X}}
\newcommand{\Y}{ {\rm Y}}
\newcommand{\Z}{{\rm Z}}
\newcommand{\Ac}{\mathcal{A}}
\newcommand{\Lc}{\mathcal{L}}
\newcommand{\Mc}{\mathcal{M}}
\newcommand{\Pc}{\mathcal{P}}
\newcommand{\Qc}{\mathcal{Q}}
\newcommand{\Tc}{\mathcal{T}}
\newcommand{\Sc}{\mathcal{S}}
\newcommand{\Bc}{\mathcal{B}}
\newcommand{\Ec}{\mathcal{E}}
\newcommand{\Fc}{\mathcal{F}}
\newcommand{\Hc}{\mathcal{H}}
\newcommand{\Uc}{\mathcal{U}}
\newcommand{\Vc}{\mathcal{V}}
\newcommand{\Wc}{\mathcal{W}}
\newcommand{\Yc}{\mathcal{Y}}
\newcommand{\Ar}{{\rm A}}
\newcommand{\Br}{{\rm B}}
\newcommand{\Ir}{{\rm I}}
\newcommand{\Fr}{{\rm F}}
\newcommand{\Jr}{{\rm J}}
\newcommand{\Mr}{{\rm M}}
\newcommand{\Or}{{\rm O}}
\newcommand{\GL}{{\rm GL}}
\newcommand{\SL}{{\rm SL}}
\newcommand{\Spr}{{\rm Sp}}
\newcommand{\Rr}{{\rm R}}
\newcommand{\Zr}{{\rm Z}}
\newcommand{\gl}{\mathfrak{gl}}
\newcommand{\middd}{{\rm mid}}
\newcommand{\ev}{{\rm ev}}
\newcommand{\Pf}{{\rm Pf}}
\newcommand{\Norm}{{\rm Norm\tss}}
\newcommand{\oa}{\mathfrak{o}}
\newcommand{\spa}{\mathfrak{sp}}
\newcommand{\osp}{\mathfrak{osp}}
\newcommand{\bgot}{\mathfrak{b}}
\newcommand{\kgot}{\mathfrak{k}}
\newcommand{\g}{\mathfrak{g}}
\newcommand{\h}{\mathfrak h}
\newcommand{\n}{\mathfrak n}
\newcommand{\z}{\mathfrak{z}}
\newcommand{\Zgot}{\mathfrak{Z}}
\newcommand{\p}{\mathfrak{p}}
\newcommand{\sll}{\mathfrak{sl}}
\newcommand{\psl}{\mathfrak{psl}}
\newcommand{\agot}{\mathfrak{a}}
\newcommand{\qdet}{ {\rm qdet}\ts}
\newcommand{\Ber}{ {\rm Ber}\ts}
\newcommand{\HC}{ {\mathcal HC}}
\newcommand{\cdet}{ {\rm cdet}}
\newcommand{\tr}{ {\rm tr}}
\newcommand{\gr}{ {\rm gr}}
\newcommand{\str}{ {\rm str}}
\newcommand{\loc}{{\rm loc}}
\newcommand{\Gr}{{\rm G}}
\newcommand{\sgn}{ {\rm sgn}\ts}
\newcommand{\ba}{\bar{a}}
\newcommand{\bb}{\bar{b}}
\newcommand{\bi}{\bar{\imath}}
\newcommand{\bj}{\bar{\jmath}}
\newcommand{\bk}{\bar{k}}
\newcommand{\bl}{\bar{l}}
\newcommand{\hb}{\mathbf{h}}
\newcommand{\Sym}{\mathfrak S}
\newcommand{\fand}{\quad\text{and}\quad}
\newcommand{\Fand}{\qquad\text{and}\qquad}
\newcommand{\For}{\qquad\text{or}\qquad}
\newcommand{\OR}{\qquad\text{or}\qquad}

\renewcommand{\theequation}{\arabic{section}.\arabic{equation}}

\newtheorem{thm}{Theorem}[section]
\newtheorem{lemma}[thm]{Lemma}
\newtheorem{prop}[thm]{Proposition}
\newtheorem{cor}[thm]{Corollary}
\newtheorem{conj}[thm]{Conjecture}
\newtheorem*{mthm}{Main Theorem}
\newtheorem*{mthma}{Theorem A}
\newtheorem*{mthmb}{Theorem B}

\newtheorem{defprop}[thm]{Definition-Proposition}
\newtheorem{defthm}[thm]{Definition-Theorem}

\theoremstyle{definition}
\newtheorem{definition}[thm]{Definition}

\newtheorem{remark}[thm]{Remark}
\newtheorem{example}[thm]{Example}

\newcommand{\bth}{\begin{thm}}
\renewcommand{\eth}{\end{thm}}
\newcommand{\bpr}{\begin{prop}}
\newcommand{\epr}{\end{prop}}
\newcommand{\ble}{\begin{lem}}
\newcommand{\ele}{\end{lem}}
\newcommand{\bco}{\begin{cor}}
\newcommand{\eco}{\end{cor}}
\newcommand{\bde}{\begin{defin}}
\newcommand{\ede}{\end{defin}}
\newcommand{\bex}{\begin{example}}
\newcommand{\eex}{\end{example}}
\newcommand{\bre}{\begin{remark}}
\newcommand{\ere}{\end{remark}}
\newcommand{\bcj}{\begin{conj}}
\newcommand{\ecj}{\end{conj}}

\newcommand{\bal}{\begin{aligned}}
\newcommand{\eal}{\end{aligned}}
\newcommand{\beq}{\begin{equation}}
\newcommand{\eeq}{\end{equation}}
\newcommand{\ben}{\begin{equation*}}
\newcommand{\een}{\end{equation*}}

\newcommand{\bpf}{\begin{proof}}
\newcommand{\epf}{\end{proof}}

\def\beql#1{\begin{equation}\label{#1}}

\newcommand{\RR}{\mathbb{R}}

\newcommand{\cO}{{\mathcal O}}
\newcommand{\cQ}{{\mathcal Q}}
\newcommand{\sC}{{C\!\!\!\!/\,}}
\newcommand{\fL}{{\mathfrak L}}
\newcommand{\fR}{{\mathfrak R}}
\newcommand{\fQ}{{\mathfrak Q}}
\newcommand{\fS}{{\mathfrak S}}
\newcommand{\fC}{{\mathfrak C}}
\newcommand{\fP}{{\mathfrak P}}
\newcommand{\fK}{{\mathfrak K}}
\newcommand{\fJ}{{\mathfrak J}}
\newcommand{\fB}{{\mathfrak B}}
\newcommand{\fF}{{\mathfrak F}}
\newcommand{\fe}{{\mathfrak e}}
\newcommand{\fh}{{\mathfrak h}}

\newcommand{\anti}{{\rm S}}
\newcommand{\tilh}{{\tilde h}}

\newcommand{\sfR}{{\mathfrak R}\!\!\!\!/\,}
\newcommand{\sfL}{{\mathfrak L}\!\!\!/}
\newcommand{\sfC}{{\mathfrak C}\!\!\!/}
\newcommand{\cT}{\mathcal T}
\newcommand{\Tr}{{\rm Tr}}
\newcommand{\bs}[1]{\boldsymbol{#1}}
\newcommand{\alg}[1]{\mathfrak{#1}}
\newcommand{\el}{\nonumber}
\newcommand{\nln}{\nonumber\\}
\newcommand{\eg}{\widetilde{{\mathfrak g}}}

\newcommand{\ms}{\medskip}

\newcommand{\cM}{{\cal M}}

\newcommand{\cp}{{\mathbb C}{\rm P}^1}

%
\newcommand{\al}{\alpha}
\newcommand{\be}{\beta}
\newcommand{\ga}{\gamma}
\newcommand{\ep}{\epsilon}
\newcommand{\de}{\delta}
\newcommand{\ka}{\kappa}
\newcommand{\la}{\lambda}
\newcommand{\ta}{\tau}
\newcommand{\dis}[1]{$\displaystyle{#1}$}
\newcommand{\bra}[1]{ \langle {#1} |}
\newcommand{\ket}[1]{ | {#1} \rangle}
\newcommand{\vev}[1]{ \langle {#1} \rangle}
\newcommand{\ol}[1]{\overline{#1}}
\newcommand{\red}[1]{\textcolor{red}{#1}}
\newcommand{\bp}{\bm{p}}
\newcommand{\bg}{\bm{g}}
\newcommand{\ee}{\bm{e}}
\newcommand{\bnu}{\bm{\nu}}
\newcommand{\bom}{\bm{\om}}
\newcommand{\bA}{\bm{A}}
\newcommand{\bB}{\bm{B}}
\newcommand{\e}{{\rm e}}

\newcommand{\im}{{\rm Im}}
\newcommand{\uu}{\mathfrak{u}}

\newcommand{\secref}[1]{\S\,\ref{#1}}
\newcommand{\appref}[1]{App.\,\ref{#1}}
\newcommand{\ssecref}[1]{\S\,\ref{#1}}
\newcommand{\figref}[1]{Fig.\,\ref{#1}}
\newcommand{\tabref}[1]{Tab.\,\ref{#1}}
\newcommand{\thmref}[1]{Thm.\,\ref{#1}}
\newcommand{\propref}[1]{Prop.\,\ref{#1}}
\newcommand{\corref}[1]{Cor.\,\ref{#1}}
\newcommand{\defref}[1]{Def.\,\ref{#1}}
\newcommand{\remref}[1]{Rem.\,\ref{#1}}
\newcommand{\lemref}[1]{Lem.\,\ref{#1}}
\newcommand{\dpropref}[1]{Def.-Prop.\,\ref{#1}}
\newcommand{\dthmref}[1]{Def.-Thm.\,\ref{#1}}



\title{\Large\bf 
Rotation angles of a rotating disc  as \\ 
the holonomy of the Hopf fibration} 
\author{Takuya Matsumoto}

\date{} 
\maketitle

\noindent
{\it
Department of Applied Physics, 
Faculty of Engineering, 
University of Fukui, 
$3$-$9$-$1$ Bunkyo, Fukui-shi, Fukui $910$-$8507$, Japan
}
\\
{\it
Department of 
Fundamental Engineering for Knowledge-Based Society, 
Graduate School of Engineering, University of Fukui, 
$3$-$9$-$1$ Bunkyo, Fukui-shi, Fukui $910$-$8507$, Japan
}
\\
~\\
E-mail: {\tt takuyama@u-fukui.ac.jp} 

\vspace{20mm}

\begin{abstract}
We address the fundamental question of how rotation angles arise 
in classical rigid body motion from a geometric viewpoint.
As a fundamental kinematical setting, we consider a rotating-disc model.
We show that the rotation angle accumulated in a cyclic motion can be naturally 
decomposed into a dynamical phase and a geometric phase, 
and that 
the latter is obtained from the $U(1)$ holonomy of the Hopf fibration 
equipped with its canonical connection
through a natural weight-two representation of the fiber group. 
By introducing a Gauss map based on a normal vector fixed 
at the center of the rotating disc, 
the physical periodic motion is mapped to 
a closed curve on the two-sphere $S^2$, 
whose horizontal lift in the Hopf bundle 
$S^3\! \to \!S^2$ determines the geometric phase
through the weight-two representation.
This framework provides a conceptually clear 
interpretation of rotation angles 
as geometric phases and clarifies the geometric origin 
of their decomposition. 
\end{abstract}


\setcounter{footnote}{0}
\setcounter{page}{0}
\thispagestyle{empty}

\newpage

\tableofcontents

\section{Introduction and summary}
\label{sec:int}
\setcounter{equation}{0}


Rotation angles arising in simple mechanical systems often exhibit features 
that cannot be fully explained by purely local or dynamical considerations. 
A classical example is provided by a disc rolling without slipping along a prescribed path, 
where the total rotation accumulated after a closed cycle generally differs from 
what would be expected from naive kinematical arguments. 
Such phenomena are naturally described in terms of a decomposition of the total rotation angle into 
a dynamical contribution and an additional phase of geometric origin. 
Similar decompositions appear in a wide range of contexts, 
from rigid body motion to geometric phases in classical and quantum mechanics.

\ms 

The purpose of the present paper is not to introduce a new mechanical model, 
but to clarify the geometric nature of this additional rotation in a particularly simple and concrete setting. 
We show that the geometric contribution to the rotation angle of a disc rolling along the rim of a fixed disc 
admits a natural interpretation as the holonomy associated with parallel transport in a principal $U(1)$ bundle. 
More precisely, by representing the kinematics of the rolling disc through an appropriate Gauss map, 
the motion is mapped to a closed curve on the two-sphere $S^{2}$, and the accumulated rotation 
is obtained from the U(1) holonomy 
of the Hopf fibration $S^{3} \to S^{2}$ 
endowed with its canonical connection 
through the weight-two representation. 
In this way, a familiar mechanical system provides a concrete realization 
of a fundamental concept in differential geometry.

\ms 


This article provides a detailed mathematical exploration of a
simple kinematical model, 
called {\it “Rotation angles of a rotating disc,”} 
in order to elucidate the concept of geometric phase, 
a universal and interdisciplinary phenomenon in physics
\cite{Pan, Lon, han, Berry1, Berry2, SW}\,.  
The primary goal is to interpret the geometric contribution 
to the rotation angle in terms of the holonomy of the Hopf
fibration and its natural weight-two representation.
This clarifies the geometric significance of its components. 
Hence, our model is a typical model that exhibits the 
geometric phase in a simple setup.

\paragraph{The physical model}
The setup involves two discs (Disc A and B) in a three-dimensional space $\mathbb{R}^3$\,.  
Disc A is a fixed disc with radius $a$\,,  
centered at the origin on the $z=0$ plane (the $xy$-plane)\,, 
while  Disc B with radius $b$ 
is another disc that rolls on the edge of Disc A without slipping. 
The position of Disc B is determined by two parameters: the angle $\theta$, 
which indicates the point of contact on Disc A's circumference, and the angle $\be$ 
between Disc B and $xy$-plane.  
The motion is described over a time interval $t \in [0, 1]$, 
during which Disc B starts from an initial configuration and returns to the same position at $t=1$. 

\paragraph{
The central question and decomposition of rotation angle} 
The core question this paper seeks to answer is as follows: 
"How much does the disc B rotate in this motion from $t = 0$ to $t = 1$?"  
The total rotation angle, denoted as $\Delta$, is shown to be decomposable into two distinct parts, 
the {\it dynamical phase} ($\Delta_d$) and {\it geometric phase} ($\Delta_g$)\,. 
The former, $\Delta_d$\,, depends on the ratio of the 
radii of Discs A and B; for a one-cycle motion, 
$\De_d=2\pi a/ b$\,. 
The latter, $\Delta_g$\,, depends only on the geometric 
motion of Disc B as a rigid body in $\RR^3$\,.
The article presents explicit formulas for both phases, 
but notes that the geometric meaning of the formula \eqref{eq:geom} 
for $\Delta_g$ is initially opaque and requires further elaboration.

\paragraph{Geometric interpretation via Gauss map}
To unravel the geometric meaning, this study introduces a {\it Gauss map}. 
This map associates the orientation of Disc B with a unit normal vector, $g(\theta, \beta)\in S^2 \subset \RR^3$. 
As Disc B completes its cyclic motion, 
the tip of this vector traces an oriented closed curve $\ga$ 
(the {\it Gauss curve}) on the unit two-sphere ($S^2$). 
The geometric phase $\Delta_g$ is then expressed 
in terms of geometric properties of this curve $\ga$\,. 
When the Gauss curve is simple, the geometric phase 
$\De_g$
is expressed in terms of geometric properties of $\ga$ by
\begin{align}
\label{eq:De-int}
\Delta_g = A_+ - 2\pi I_+\,, 
\end{align}
where $A_+$ is the area enclosed by the curve on the sphere,
and $I_+$ is a {\it topological number} representing 
the number of poles enclosed by $\ga(\ep)$\,, 
a regularized version of the curve. 

\paragraph{The Hopf fibration and holonomy} 
The main theoretical development of the paper compared 
with our previous work \cite{MTY} 
is to connect this geometric phase to the underlying mathematical structure, the {\it Hopf fibration}. 
The Hopf fibration describes the three-sphere 
($S^3$) as a principal fiber bundle over 
$S^2$ with a $U(1)$ fiber.  
The article demonstrates that the physical geometric phase 
is obtained from the $U(1)$ {\it holonomy} $\tau(\ga)$ 
of the Hopf fibration associated with the Gauss curve $\ga$,
through the natural weight-two representation 
of the fiber group.
The key ingredient is the {\it canonical connection} 
$\om$ in \dpropref{def:cc}\,, 
which determines the horizontal lift of the Gauss curve.  
The main theorem of this study is as follows. 
\ms 
\begin{mthm}[\thmref{thm:hol}]
Let $\ga$ be a piecewise smooth Gauss curve on 
$\cp$ associated with 
the motion \eqref{eq:motion}\,. 
Let
\begin{align}
\label{eq:rho2-int}
\rho _2:U(1)\longrightarrow U(1),
\qquad
\rho _2(u)=u^2,
\end{align}
be the weight-two representation of the fiber group. 
Then the geometric phase $\De_g$ and the holonomy 
$\tau(\ga)$ of the Hopf fibration are related by
\begin{align}
\rho _2(\tau(\ga))=\exp(i\De_g).
\label{eq:int-hol}
\end{align}
\end{mthm}

\ms 

Thus, the physical geometric phase is obtained from the
$U(1)$ holonomy of the Hopf fibration through the
natural weight-two representation $\rho _2$. 
This relation provides a precise geometric interpretation 
of the rotation angle that is not apparent from the original
kinematical formula \eqref{eq:geom}.

\ms 


On the other hand, the $U(1)$ relation \eqref{eq:int-hol} 
loses the real-valued information contained in the geometric
phase, including the topological contribution appearing in 
\eqref{eq:De-int}. 
This issue is addressed in \secref{sec:cover}
by considering a real lift of the $U(1)$ holonomy. 
Let $p:\RR\to U(1)$, $p(x)=e^{ix}$ 
be the universal covering map, and let
\begin{align}
\label{eq:rhotil2-int}
\wt{\rho}_2:\RR\longrightarrow\RR,
\qquad
\wt{\rho}_2(x)=2x,
\end{align}
be the lift of the weight-two representation $\rho_2$. 
For the representative of the lifted holonomy 
$\wt{\tau}_M(\ga)$
associated with the motion \eqref{eq:motion}, 
we obtain $\wt{\tau}_M(\ga)
=\pi n+\Delta_g/2$. 
Consequently,
\begin{align}
\wt{\rho}_2\left(\wt{\tau}_M(\ga)\right)
=2\pi n+\Delta_g\,.
\end{align}
These relations are summarized by the commutative diagram
\begin{align}
\begin{CD}
\RR\ni\wt{\tau}_M(\ga)
@>{\wt{\rho}_2}>\eqref{eq:rhotil2-int}>
\RR\ni 2\pi n+\De_g
\\
@V{p}VV @VV{p}V
\\
U(1)\ni\tau(\ga)
@>{\rho_2}>\eqref{eq:rho2-int}>
U(1)\ni \exp({i\De_g})\,. 
\end{CD}
\end{align}
Thus, the passage from the Hopf holonomy to the physical
geometric phase can be described consistently 
both at the $U(1)$ level and at the level of 
its universal covering space.

\ms 

A complementary geometric interpretation is developed in 
\secref{ssec:geom-curv}. 
Let $\phi$ denote the angle of the tangent vector of 
the Gauss curve relative to a local moving frame on $S^2$, 
and let $\ka_g$ be its geodesic curvature. 
Then, the geometric phase $\De_g(t)$ for any $t\in [0,1]$
is determined by 
the horizontal condition for the canonical connection 
and given by (\propref{prop:De-t}) 
\begin{align}
\De_g(t)
=\phi(t)-\phi(0)-\int_0^{s(t)}\ka_g(s)\,ds ,
\end{align}
where $s$ is the length parameter. 
Hence, the geometric phase can also be understood directly 
on the two-sphere as a combination of 
the net rotation of the tangent direction relative 
to the local frame and 
the accumulated geodesic curvature.

\paragraph{Contents} 
The remainder of this paper is organized as  follows. 
In \secref{sec:def}, we set up the model called  
{\it Rotation angles of a rotating disc.}   
\secref{sec:hopf} is devoted to 
collecting the basic materials on the Hopf  fibration, including 
the definition (\secref{sec:def-hopf}), 
the canonical connection (\secref{sec:conn}),
the decomposition of $S^3$ into the horizontal and 
vertical subspaces (\secref{sec:hori-ver}),  
and the notion of parallel displacement (\secref{sec:para}). 
In \secref{sec:rot-hol}, we demonstrate that 
the geometric phase of the rotation angle can 
be obtained from the $U(1)$ holonomy of the Hopf fibration
through the weight-two representation (\secref{sec:hol}). 
We then introduce a real lift of the $U(1)$ holonomy and 
clarify how the real-valued geometric phase, 
including its topological information, 
is retained at the level of the universal covering space 
$\RR$ (\secref{sec:cover}). 
Finally, we relate the canonical Hopf connection to 
the local moving frame and the geodesic curvature 
of the Gauss curve, providing a complementary 
geometric interpretation of the geometric phase 
directly on $S^2$ (\secref{ssec:geom-curv}).
In \appref{app:diffeo}, we explicitly present a 
diffeomorphism of $S^2$ and $\cp$.


\section{The model: Rotation angles of a rotating disc}
\label{sec:def}
\setcounter{equation}{0}

In this section, we introduce a simple kinematical model\footnote{
This model was initially proposed as one of the
common problems 
for the 32nd Japan Mathematics Contest and 
the 25th Japan Junior Mathematics Contest \cite{JMC}.}, 
which we refer to as 
{\it ``Rotation angles of a rotating disc,''}   
and summarize some of the results obtained for the model in \cite{MTY}\,. 

\ms 

\begin{itembox}[l]{\bf \large 
Rotation angles of a rotating disc}
Consider a disc A of radius $a>0$ lying in $z=0$ plane 
({\it i.e.} $xy$-plane) 
with its center at the origin
in three-dimensional space $\RR^3$\,. 
Set another disc B in $\RR^3$ with the radius $b>0$
so that it contacts disc A at 
$(a\cos\theta, a\sin \theta,0)\in \RR^3$ with 
$\theta \in \RR$\,. 
Suppose that the line perpendicular to disc B and 
passing through its center is either parallel to the $z$-axis 
or intersects the $z$-axis.
Denote the angle between $z=0$ plane and the disc B by $\beta$\footnote{
Comparing with our previous work \cite{MTY}\,, 
the definition of $\beta$ is changed as 
$\beta_{\rm [here]}=\pi-\beta_{\cite{MTY}}$\,. }\,, 
and suppose that $0\leq \beta \leq\pi$\,, 
see Fig.\ref{fig:prob}\,. 
The position of disc B is determined entirely by two parameters, $\theta$ and $\be$\,.  

\ms 

Let disc B roll along the edge of disc A without slipping.

\ms

Introduce the time parameter $t\in [0,1]\subset \RR$
to describe this motion,
and regard the angle variables 
$\theta$ and $\be$ as continuous functions of $t$\,.  
Then, the position of the rigid body, disc B, 
is determined by the map 
\begin{align}
(\theta, \beta): [0,1]\to \RR \times [0, \pi]\,, 
\quad 
t\mapsto (\theta(t), \beta(t))\,. 
\label{eq:motion}
\tag{M}
\end{align}
Set $\theta(0)=0$\,. 
Starting at $t=0$, disc B returns to its initial position at $t=1$\,. 
That is 
\begin{align}
\theta(1)=2\pi n
\quad \text{with} \quad n\in \ZZ\,, 
\quad \be(1)=\be(0)\,. 
\tag{T}
\label{eq:top}
\end{align}
We shall call the integer $\theta(1)/2\pi=n$ the 
{\it topological number}. 


\end{itembox}

\begin{figure}[htbp]
\centering
\includegraphics[keepaspectratio, scale=0.4]{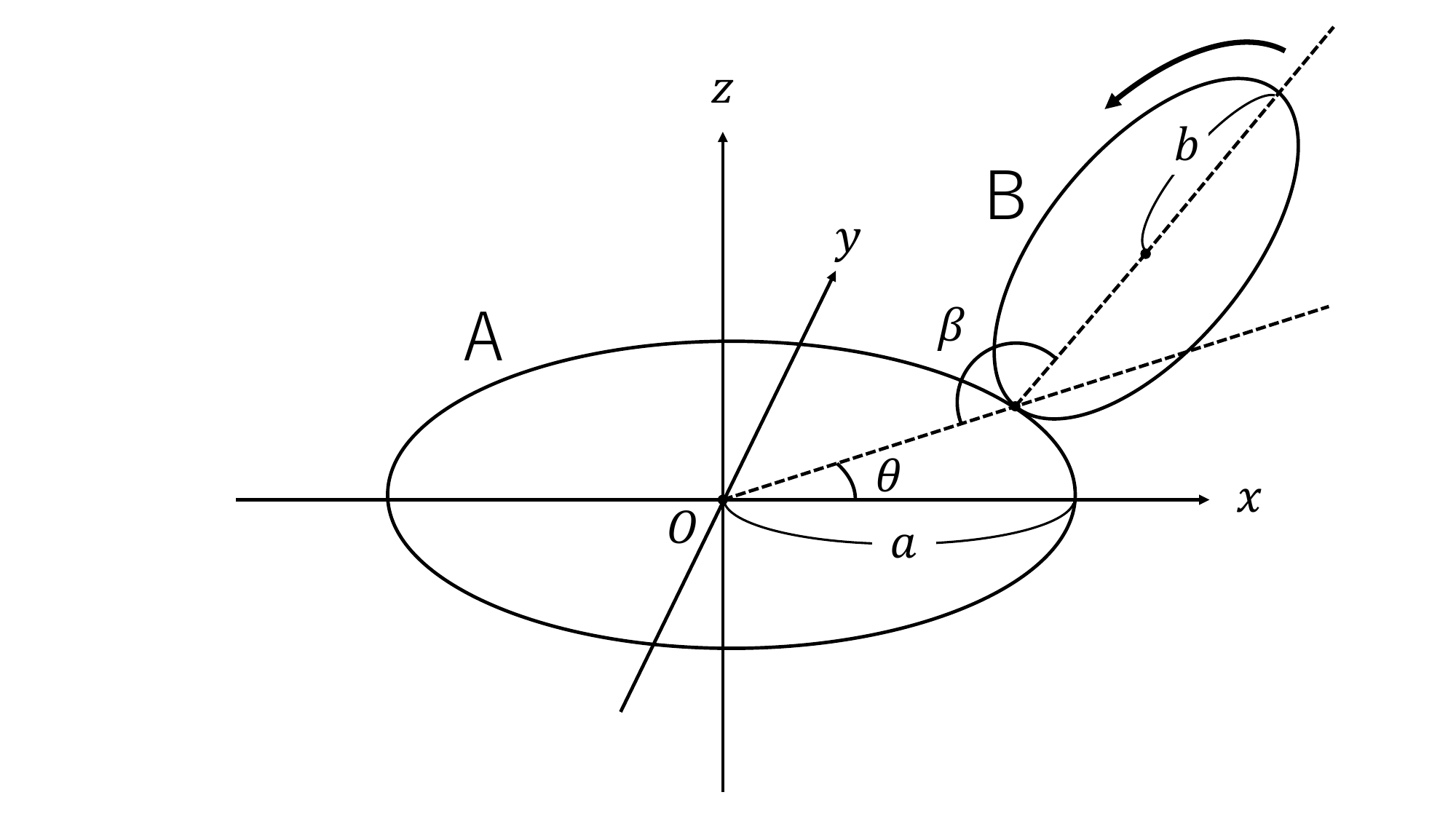}
\caption{Disc A is fixed on $xy$-plane and 
disc B 
rolls on the edge of the disc A 
without slipping. }
\label{fig:prob}
\end{figure}

\ms 

\begin{remark}
\label{rem:lip}
The functions $\theta(t)$ and $\beta(t)$
are assumed to satisfy the following conditions.
\begin{enumerate}
\item The continuous functions $\theta(t)\,, \beta(t)$  
satisfy the {\it Lipschitz condition}. 
That is, for any $t_1\,, t_2\in [0,1]$\,, there exists a positive 
constant $C_f$ such that 
\begin{align}
\left|f(t_1)-f(t_2)\right|\leq C_f |t_1-t_2| \,,
\end{align}
where $f$ is either $\theta$ or $\be$\,. 

\item Both $\theta(t)$ and $\beta(t)$ are differentiable 
except at finitely many points on $(0,1)$.  
The derivatives $\theta'(t)\,, \be'(t)$ are 
piecewise continuous on $[0,1]$\,.  

\end{enumerate}  
These two conditions are physically natural. 
Since $t\in [0,1]$ is an auxiliary parameter to describe the motion 
\eqref{eq:motion}\,, 
it is always possible to assume that  $\theta(t)\,, \be(t)$ 
satisfy the above conditions. 
\end{remark}

\ms 

Our interest lies in the following question. 

\ms 
\begin{itembox}[l]{\bf Question}
How much does the disc B rotate in this motion from 
$t=0$ to $t=1$? 
Define and determine the rotation angle $\De$\,. 
\end{itembox}
\ms 

The answer to this question was provided by \cite{MTY}\,. 

\begin{defthm}[\cite{MTY}]
\label{thm:line-int}
The total rotation angle $\De$ of disc B is decomposed into the sum of 
the dynamical phase $\De_d$ and the geometric phase $\De_g$\,, 
\begin{align}
\De=\De_d+\De_g\,, 
\end{align}
and they are respectively given by 
\begin{align}
\De_d&=\frac{2\pi n a}{b}\qquad && \text{\rm (Dynamical phase)}\,, \\ 
\De_g&=-\int_{0}^{1} \cos\be(t) \frac{d\theta(t)}{dt} dt\qquad && \text{\rm (Geometric phase)}\,. 
\label{eq:geom}
\end{align}
\end{defthm}

\ms 

Though the above theorem gives us an explicit formula
for the geometric phase, the geometric meaning is opaque. 
We elaborate on this point in the following subsections.  

\ms

\subsection*{Gauss map}

First, we introduce the unit normal vector of disc B by {\it the Gauss map}, 
which detects the motion of disc B. 
We assume that two-sphere $S^2$ is 
embedded in $\RR^3$ as 
\begin{align}
S^2=\{~x^2+y^2+z^2=1~|~\begin{pmatrix} x\\ y\\ z  \end{pmatrix}
\in \RR^3~\}\subset \RR^3 \,. 
\notag 
\end{align}
\begin{definition}[Gauss map]
\label{def:gmap}
For the two parameters $(\theta, \beta)$ in the motion 
\eqref{eq:motion}\,, 
define {\it the Gauss map} from $\RR \times [0,\pi]$
to two-sphere $S^2$ by
\begin{align}
\bg :\RR \times [0,\pi]\to S^2\subset \RR^3
\,, \quad 
(\theta, \beta)
\mapsto 
\bg(\theta, \beta)=\begin{pmatrix}
\sin\be \cos\theta \\ 
\sin\be \sin\theta \\ 
\cos\be
\end{pmatrix} \,. 
\label{eq:gauss_a}
\end{align}
The vector $\bg(\theta, \beta)$ is called {\it the Gauss vector}. 
\end{definition}

The Gauss vector is geometrically realized both in the rotating model
and two-sphere as in \figref{fig:Gmap-1} and 
\figref{fig:Gmap-2}\,, respectively. 

\ms 

\begin{figure}
\centering
\begin{subfigure}{0.45\columnwidth}
\centering
\includegraphics[width=1.3\columnwidth]{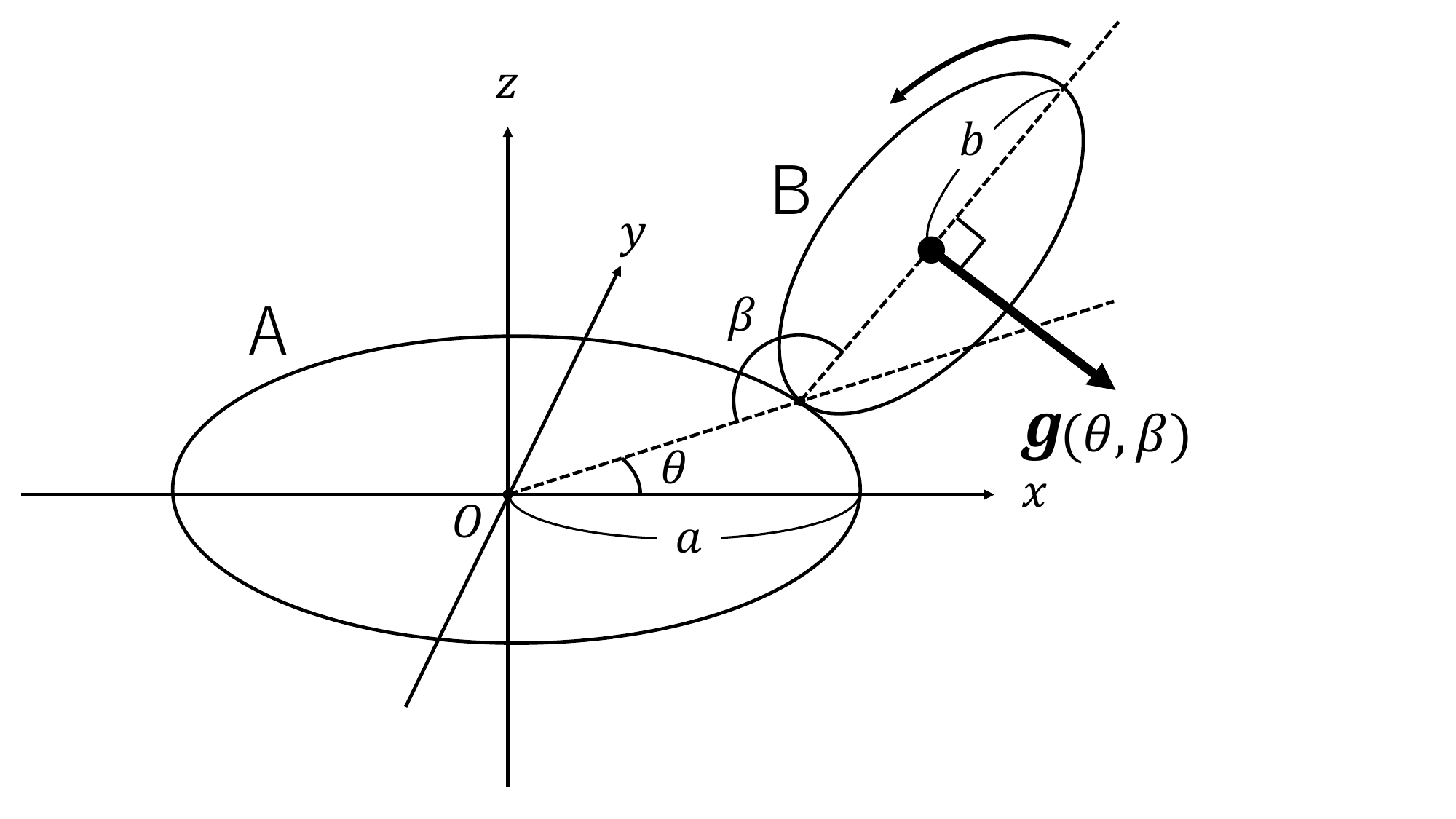}
\caption{The Gauss vector  $\bg$  in the model}
\label{fig:Gmap-1}
\end{subfigure}
\hspace{3mm}
\begin{subfigure}{0.45\columnwidth}
\centering
\includegraphics[width=1.4\columnwidth]{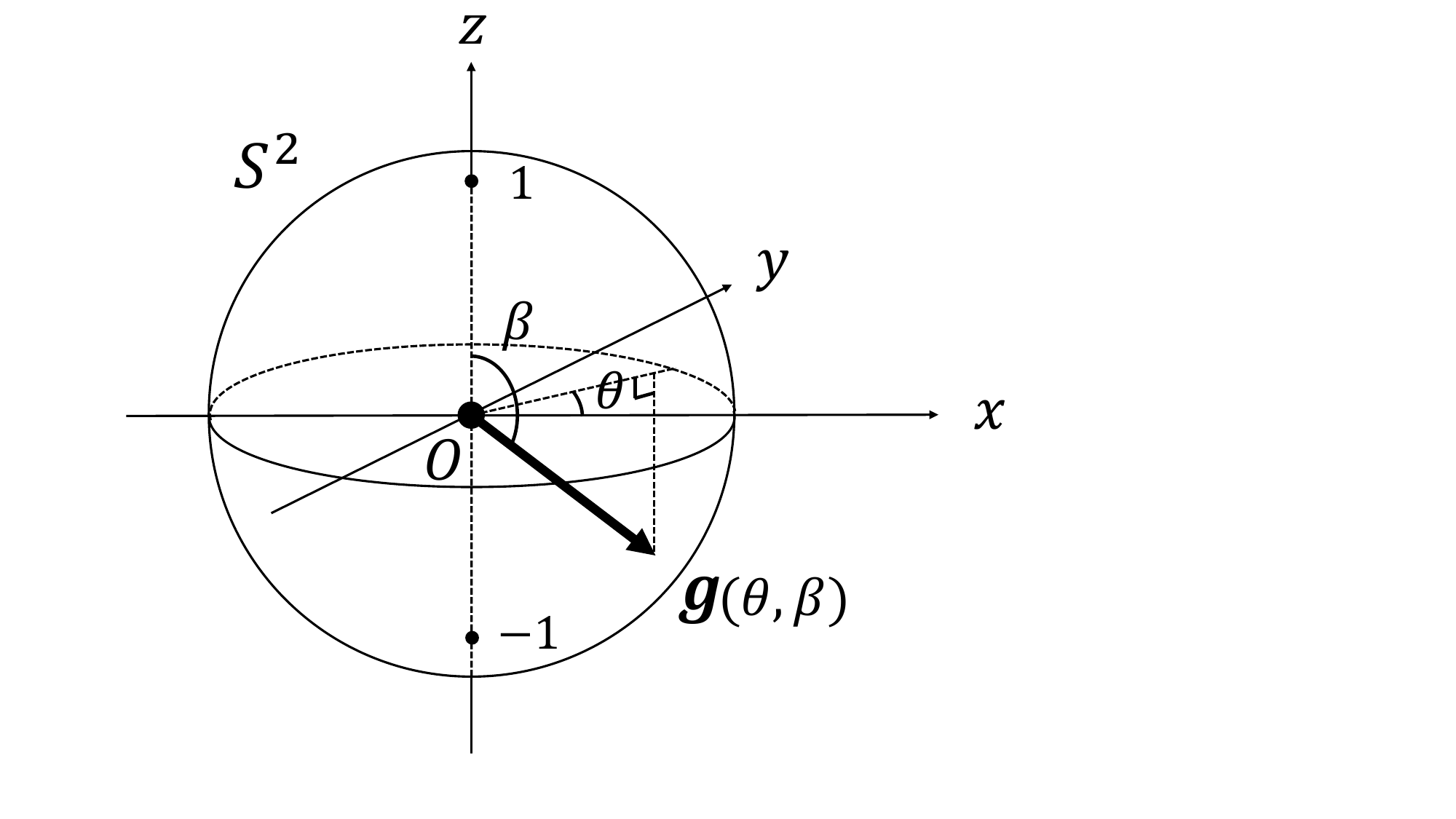}
\caption{The Gauss vector $\bg$ in two-sphere}
\label{fig:Gmap-2}
\end{subfigure}
\caption{The Gauss map $\bg$ from the model (a) 
to two-sphere (b)}
\label{fig:Gmap}
\end{figure}

Note that $\bg(\theta, \beta)$ is a unit vector and 
perpendicular to disc B.    
By composing the motion \eqref{eq:motion} and the Gauss
map \eqref{eq:gauss_a}, the Gauss vector could be regarded
as a function of $t\in [0,1]$\,, 
\begin{align}
\bg : [0,1] \to S^2\,, \quad t\mapsto \bg(t)
=\begin{pmatrix}
\sin\be(t) \cos\theta(t) \\ 
\sin\be(t) \sin\theta(t) \\ 
\cos\be(t)
\end{pmatrix} \,. 
\label{eq:gauss}
\end{align}
We also refer to the map in \eqref{eq:gauss} as the Gauss map. 

\ms 

Because of $\bg(1)=\bg(0)$\,, the Gauss map defines 
the oriented closed curve $\ga$ on $S^2$ 
associated with 
the given motion \eqref{eq:motion} by 
\begin{align}
\ga=\{~\bg(t)\in S^2~|~t\in [0,1]~\}\subset S^2\,.  
\label{eq:ga}
\end{align}
The orientation of $\ga$ is induced from that of 
the parameter $t\in [0,1]$ such as from $t=0$ to $t=1$\,.  
The curve $\ga$ is said to be {\it simple} if it does not intersect with itself.

\subsection*{The regularized curve}
Second, though the Gauss map \eqref{eq:gauss} detects the motion of the rotating disc, 
it does not move when the disc is in the $xy$ plane, which corresponds to  $\be=0, \pi$\,. 
To resolve this incompleteness, we introduce the following regularization.  
For a sufficiently small $\ep\in (0, \pi/8)$\,, which plays the role of {\it cut-off}\,, 
define the regularized angular coordinate by 
\begin{align}
&\be_\ep: [0,1] \to [\ep \,, \pi-\ep ] \,, 
\qquad t\mapsto \be_\ep(t) \,, 
\nln 
&\be_\ep(t)=\begin{cases}
\ep & (~0\leq \be(t) \leq \ep~) \\
\be(t) &(~\ep < \be(t) <\pi-\ep~) \\ 
\pi-\ep & (~\pi- \ep \leq \be(t) \leq \pi~) \,. 
\end{cases}
\end{align}
The regularized Gauss vector is then given by replacing $\be(t)$
by $\be_\ep(t)$ as  
\begin{align}
\bg_\ep : [0,1] \to S^2-\{0,0,\pm1\}\,, \quad 
t\mapsto \bg_\ep(t)
=\begin{pmatrix}
\sin\be_\ep(t) \cos\theta(t) \\ 
\sin\be_\ep(t) \sin\theta(t) \\ 
\cos\be_\ep(t)
\end{pmatrix} \,. 
\label{eq:gauss-ep}
\end{align}

\ms 

\begin{definition}[Regularized Gauss curve]
\label{def:ga-reg}
The {\it regularized Gauss curve} is defined by the orbit
of the regularized Gauss vector,  
\begin{align}
\ga(\ep)=\{~\bg_\ep(t)\in S^2-\{(0,0,\pm1)\}~
|~t\in [0,1]~\}\subset 
S^2-\{(0,0,\pm1)\}\,.  
\label{eq:ga-reg}
\end{align}
\end{definition}

\ms 

The point is  that the curve $\ga(\ep)$ dodges the poles
$(0,0,\pm1)$\,, though the undeformed curve 
$\ga$ does not in general.  
By taking $\ep \to 0$\,, it continuously reduces to the original Gauss curve,
\begin{align}
\ga(\ep) \to \ga\,. 
\end{align}
The significance of $\ga(\ep)$ is that it correctly reflects the topological indices 
$I_\pm$ as we will see later.\footnote{
For the details, see Remark 3.7 in \cite{MTY}\,.}  
Adopting this regularization, we could rewrite the geometric phase 
\eqref{eq:geom}
in terms of the line integral, as follows: 

\ms 

\begin{prop}[\cite{MTY}]
\label{prop:gp-reg}
The geometric phase for the motion \eqref{eq:motion}
is given by the limit of the line integral along the regularized curve
$\ga(\ep)$ in \eqref{eq:ga-reg}\,, 
\begin{align}
\De_g
=-\lim_{\ep\to0}\int_{\ga(\ep)} \cos\be_\ep d\theta
\label{eq:gp-reg}
\end{align}
\end{prop}

\ms 

Furthermore, the geometric meaning of the above line integral has been 
revealed in \cite{MTY}\,.  
Denote by $A_+$ $(A_-)$ the area enclosed by the Gauss curve $\ga$ on the left (right), 
and by $I_+ (I_-)\in \{0,1,2\}$ the total number of poles at $(0,0,\pm1)\in S^2$  
enclosed by the regularized Gauss curve $\ga(\ep)$ on the left (right, respectively). 
By the definitions, they satisfy 
\begin{align}
\label{eq:Apm}
A_++A_-=4\pi \qquad\text{and}\qquad  I_++I_-=2\,. 
\end{align}
With these notations, the geometric phase $\De_g$ is expressed as follows: 

\ms 

\begin{thm}[Theorem 3.15 in \cite{MTY}]
Suppose that the oriented closed curve $\ga$ defined in 
\eqref{eq:ga} is simple.
Then, for the motion \eqref{eq:motion}\,, 
the geometric phase is given by 
\begin{align}
\De_g
&=A_+-2\pi I_+
=-A_-+2\pi I_- 
=\frac{A_+-A_-}{2}-\pi(I_+-I_-) \,. 
\label{eq:mainthm}
\end{align}
\end{thm}

\ms 

Together with the dynamical phase $\De_d$\,, 
the answer to this question is summarized as follows. 

\ms

\begin{itembox}[l]{\bf Answer for the Question}
The rotation angle $\De$ of the rotating disc B in the motion 
\eqref{eq:motion} with the topological condition \eqref{eq:top}
is given by the sum of the dynamical phase $\De_d$
and the geometric phase $\De_g$ as follows, 
\begin{align}
\De=\De_d+\De_g  \quad \text{with}\quad 
\De_d=\frac{2\pi n a}{b} 
\,,\quad 
\De_g=A_+-2\pi I_+\,.
\label{eq:ans}
\end{align}
Here, $A_+$ is the area of the region surrounded by the curve 
$\ga$ in \eqref{eq:ga} which is supposed to be simple, 
and $I_+\in \{0,1,2\}$ is the number of poles
enclosed by the regularized curve $\ga(\ep)$ in 
\eqref{eq:ga-reg}
on the left side. See also \figref{fig:mainthm}\,. 
\end{itembox}

\begin{figure}[h]
\centering
\includegraphics[width=15cm]{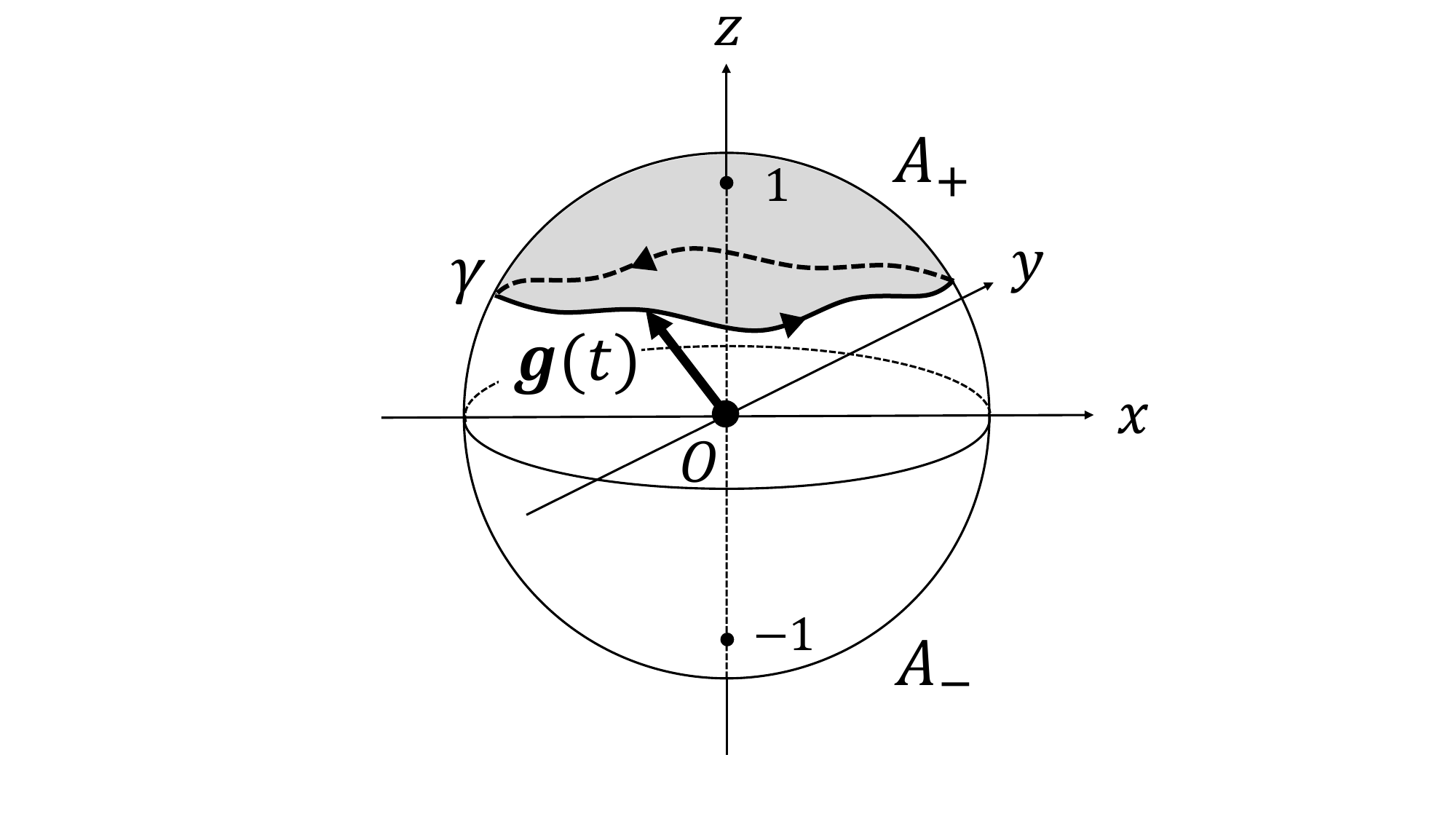}
\caption{The curve $\ga\subset S^2$ is the orbit of 
the Gauss vector $\bg(t)$ whose orientation is induced 
by the parameter $t$\,. 
$A_+ (A_-)$ is the area enclosed by $\ga$ on the left (right) side of the figure.  
The topological index $I_+ (I_-)$ is the number of poles 
at $(0,0,\pm1)\in S^2$ 
contained in $\ga$ on the left (right) side. 
The above picture corresponds to the case $I_+=I_-=1$\,. }
\label{fig:mainthm}
\end{figure}%

\section{Hopf fibration}
\label{sec:hopf}
\setcounter{equation}{0}

In this section, we shall review the definition of the Hopf fibration in \secref{sec:def-hopf} 
and fix some conventions for subsequent arguments.  
In particular, the {\it canonical connection} in \secref{sec:conn} is important 
because it will be used to introduce the {\it horizontal} and {\it vertical} subspaces 
in \secref{sec:hori-ver}\,, and define the {parallel displacement} in \secref{sec:para}\,. 
The main references in this section are \cite{KN,K,EGH,hopf}\,. 

\subsection{Definition of the Hopf fibration}
\label{sec:def-hopf}

The {\it Hopf fibration} \cite{hopf} is a principal fiber bundle equipped with  
$U(1)$ fiber embedded in the total space $S^3$
and the projection called the {\it Hopf map}
$\pi: S^3 \to S^2$\,, {\it i.e.}\,,  
\begin{align}
S^1 \hookrightarrow S^3 \xrightarrow{\pi} S^2\,. 
\end{align}

\ms 

To be more precise, define the three-sphere $S^3$ by 
\begin{align}
S^3= \{~(z_1, z_2) \in \CC^2~|~|z_1|^2+|z_2|^2=1~\}\subset \CC^2\,. 
\end{align}
The one-dimensional complex projective space
$\cp$ is a quotient space 
of $\CC^2\backslash \{(0,0)\}$ by $\CC^\times=\CC\backslash \{0\}$ action. 
Namely, the equivalence relation $(z_1, z_2)\sim (w_1, w_2)$ is defined by 
\begin{align}
(z_1, z_2)=(\la w_1, \la w_2) \quad \text{with}\quad \la\in \CC^\times\,.  
\end{align}
We denote a representative of $\cp$ by $[z_1,z_2]$\,, that is, 
\begin{align}
\cp= \{~[z_1, z_2] ~|~(z_1, z_2)\in \CC^2\backslash
\{(0,0)\}~\}\,.
\end{align}
The projective space $\cp$ has an atlas $\{(U_1, f_1), (U_2, f_2)\}$ defined by 
\begin{align}
f_1 : U_1&=\{~[z_1, z_2] ~|~z_1\neq 0~\}\to \CC\,,  
\quad f_1([z_1, z_2])=\frac{z_2}{z_1} \,, \notag \\
f_2 : U_2&=\{~[z_1, z_2] ~|~z_2\neq 0~\}\to \CC\,,  
\quad f_2([z_1, z_2])=\frac{z_1}{z_2} \,.  
\end{align}
Since $f_1(U_1\cap U_2)=f_2(U_1\cap U_2)=\CC^\times$\,, the transition map reads 
\begin{align}
f_2\circ f_1^{-1} : \CC^\times\to \CC^\times\,, 
\quad z \mapsto \frac{1}{z}\,. 
\end{align}
Note that $\cp$ and $S^2$ are diffeomorphic and the explicit correspondence 
is summarized in \appref{app:diffeo}. 

\ms

\begin{definition}[Hopf fibration]
The {\it Hopf fibration} $(S^3, \cp, \pi, U(1))$ is a principal fiber bundle $S^3$ over the base manifold $\cp\simeq S^2$ with the structure group $U(1)\simeq S^1$\,.  
They satisfy the following definition: 
\begin{itemize}
\item[(i)]
The projection $\pi$ called the {\it Hopf map} is defined by 
\begin{align}
\pi : S^3 \subset \CC^2 \to \cp\,, \quad (z_1,z_2) \mapsto [z_1,z_2] \,. 
\end{align}

\item[(ii)]
The Lie group $U(1)$ acts on $S^3$ from the right as 
\begin{align}
S^3\times U(1) \to S^3\,, \quad ((z_1,z_2),\, s)\mapsto (z_1,z_2)s=(z_1s, z_2s)\,. 
\end{align}
The $U(1)$ action on the fiber $\pi^{-1}([z_1,z_2])\subset S^3$ is {\it simply transitive}, which means the following two properties:  
\begin{enumerate}
\item[(ii a)] For any $(z_1,z_2)\in S^3$ and $s\in U(1)$\,, 
it satisfies $\pi((z_1,z_2)s)=\pi((z_1,z_2))$\,. 
\item[(ii b)] If $\pi((z_1,z_2))=\pi((z'_1,z'_2))$\,, 
then there exists a unique $s\in U(1)$ such that 
\begin{align}
(z'_1,z'_2)=(z_1,z_2)s\,. 
\end{align}
\end{enumerate}
\item[(iii)]
The {\it section} of the Hopf fibration is defined by 
\begin{align}
\si_1 : U_1 \to \pi^{-1}(U_1)\,, \quad 
\si_1([z_1, z_2])= \frac{(z_1, z_2)|z_1|/z_1}{\sqrt{|z_1|^2+|z_2|^2}} \,, \notag \\
\si_2 : U_2 \to \pi^{-1}(U_2)\,, \quad 
\si_2([z_1, z_2])= \frac{(z_1, z_2)|z_2|/z_2}{\sqrt{|z_1|^2+|z_2|^2}} \,. 
\end{align}
\end{itemize}
\end{definition}

\ms 

By the definition, the {\it transition function} can be read as 
\begin{align}
\psi_{12}: U_1\cap U_2 \to U(1)\,, \quad 
\psi_{12}([z_1, z_2])= \frac{z_1}{|z_1|} 
\left(\frac{z_2}{|z_2|}\right)^{-1} 
\label{eq:trans}
\end{align}
and it satisfies,  on $U_1\cap U_2$\,,  
\begin{align}
\psi_{12}([z_1, z_2])\psi_{21}([z_1, z_2])=1\,, 
\qquad 
\si_2([z_1, z_2])=\si_1([z_1, z_2])\psi_{12}([z_1, z_2])\,.  
\end{align}
The total space $S^3$ of the Hopf fibration could be locally regarded as a direct product of the base manifold $\cp$ and the fiber $U(1)$ due to the {\it local trivialization} given by 
\begin{align}
\phi_1 : \pi^{-1}(U_1)\to U_1\times U(1)\,, \quad 
\phi_1((z_1,z_2))=
(\,[z_1, z_2]\,, \frac{z_1}{|z_1|}\,) \,, \notag \\
\phi_2 : \pi^{-1}(U_2)\to U_2\times U(1)\,, \quad 
\phi_2((z_1,z_2))=
(\,[z_1, z_2]\,, \frac{z_2}{|z_2|}\,)\,. 
\end{align}

\subsection{Canonical connection}
\label{sec:conn}

Suppose that the hermitian inner product on $\CC^2$ is defined by
\begin{align}
\vev{(z_1, z_2)\,, (w_1, w_2)}=\ol{z_1}w_1+\ol{ z_2}w_2
\quad \text{for}\quad 
(z_1, z_2)\,, (w_1, w_2) \in \CC^2\,, 
\end{align}
where $\ol{z}$ stands for the complex conjugate of $z\in \CC$\,. 

\ms 

\begin{definition}
The {\it canonical connection} (or Berry-Simon connection) 
$\om_i$ $(i=1,2)$ is a $\alg{u}(1)=i\RR$ valued one-form on $U_i\subset \cp$ defined by 
\begin{align}
\om_1([z_1, z_2])&=\vev{\si_1([z_1, z_2])\,, d\si_1([z_1, z_2])}\,, \notag \\
\om_2([z_1, z_2])&=\vev{\si_2([z_1, z_2])\,, d\si_2([z_1, z_2])}\,. 
\end{align}
\end{definition}

\ms 

They  are explicitly given by 
\begin{align}
\om_1([z_1, z_2])&=\frac{|z_2|^2}{|z_1|^2+|z_2|^2}
\left[\frac{|z_2|}{z_2}d\left(\frac{z_2}{|z_2|}\right)
-\frac{|z_1|}{z_1}d\left(\frac{z_1}{|z_1|}\right) \right]\,, 
\notag \\
\om_2([z_1, z_2])&=\frac{-|z_1|^2}{|z_1|^2+|z_2|^2}
\left[\frac{|z_2|}{z_2}d\left(\frac{z_2}{|z_2|}\right)
-\frac{|z_1|}{z_1}d\left(\frac{z_1}{|z_1|}\right) \right]\,.
\end{align}

\ms 

\begin{prop}
The canonical connections
(Berry-Simon connections) $\om_i$ ($i=1,2$)
are mutually related by the 
{\it gauge transformation}
with the transition function \eqref{eq:trans} on $U_1\cap U_2$ as 
\begin{align}
\om_2([z_1, z_2])&=\om_1([z_1, z_2])
+\psi_{12}([z_1, z_2])^{-1}d\psi_{12}([z_1, z_2])\,, 
\notag \\
\om_1([z_1, z_2])&=\om_2([z_1, z_2])
+\psi_{21}([z_1, z_2])^{-1}d\psi_{21}([z_1, z_2])\,. 
\end{align}
\end{prop}

\begin{proof}
Since the second relation follows from the first one, it is sufficient 
to show the first relation. By the definition of the transition function
\eqref{eq:trans}\,, we have 
\begin{align}
\psi_{12}([z_1, z_2])^{-1}d\psi_{12}([z_1, z_2])
&=d\log\psi_{12}([z_1, z_2]) 
=\frac{|z_1|}{z_1} d\left(\frac{z_1}{|z_1|}\right)
-\frac{|z_2|}{z_2} d\left(\frac{z_2}{|z_2|}\right)
\notag \\
&=\om_2([z_1, z_2])-\om_1([z_1, z_2])\,. 
\end{align}
This is the desired relation. 
\end{proof}

\ms 
In the above proof, we see that the following equality holds 
on $\pi^{-1}(U_1\cap U_2)$
\begin{align}
\om_1([z_1, z_2])+
\frac{|z_1|}{z_1} d\left(\frac{z_1}{|z_1|}\right)
=
\om_2([z_1, z_2])+
\frac{|z_2|}{z_2}d\left(\frac{z_2}{|z_2|}\right)\,. 
\end{align}
In other words, introducing $\alg{u}(1)$ valued one-form 
on $\pi^{-1}(U_i)\simeq U_i\times U(1)$ by 
\begin{align}
\wt{\om}_i((z_1, z_2))=\om_i([z_1, z_2])+
\frac{|z_i|}{z_i} d\left(\frac{z_i}{|z_i|}\right)\,, 
\qquad 
(i=1,2)
\end{align}
we obtain
\begin{align}
\wt{\om}_1((z_1, z_2))=\wt{\om}_2((z_1, z_2))
\quad \text{on}\quad 
\pi^{-1}(U_1\cap U_2)\,. 
\end{align}
Thus, the $\alg{u}(1)$ valued one-form $\om$ given 
as below is globally defined on $\cp$\,. 

\ms 
\begin{defprop}[Canonical connection]
\label{def:cc}
Define $\alg{u}(1)$ valued one-form on $\pi^{-1}(U_i)\subset S^3$ 
$(i=1,2)$ by 
\begin{align}
\om((z_1, z_2))&=\om_i([z_1, z_2])+
\frac{|z_i|}{z_i} d\left(\frac{z_i}{|z_i|}\right)\,. 
\end{align}
Then, one-form $\om$ is 
globally defined on $S^3$\,. 
Explicitly, it is given by 
\begin{align}
\om((z_1, z_2))
&=\vev{(z_1, z_2), d(z_1, z_2)}
\notag \\
&=\frac{1}{2}\left(\ol{z}_1dz_1+\ol{z}_2dz_2
-z_1d\ol{z}_1-z_2d\ol{z}_2\right)
\qquad \text{for}\qquad (z_1,z_2)\in S^3\,. 
\label{eq:cc}
\end{align}
We refer to $\om$ as the canonical connection (or Berry-Simon connection) on $S^3$\,. 
\end{defprop}

Note that the second equality in \eqref{eq:cc} follows by 
differentiating $\vev{(z_1,z_2), (z_1,z_2)}=1$\,. 
Schematically, the connections $\om$ and $\om_i$ $(i=1,2)$ are related as
\begin{align}
\om=\pi^*(\om_i)\qquad \text{and} \qquad 
\si_i^*(\om)=\om_i \,. 
\end{align}

\subsection{Horizontal and vertical subspaces}
\label{sec:hori-ver}

The connection one-form $\om$ defined in \eqref{eq:cc} satisfies two important properties, as stated below. 
For an element of Lie algebra $A\in \alg{u}(1)$\,, consider the one-parameter subgroup 
\begin{align}
\{~e^{tA}~|~t\in \RR~\}\subset S^3\,. 
\end{align}
Let us make this subgroup act on $(z_1,z_2)\in S^3$ from right and define the vector 
$(A^*)_{(z_1,z_2)}\in T_{(z_1,z_2)}S^3$ as the tangent vector of the orbit $(z_1,z_2)e^{tA}$ at $t=0$\,, 
\begin{align}
(A^*)_{(z_1,z_2)}=\frac{d}{dt}\left((z_1,z_2)e^{tA}\right)\big|_{t=0}=(z_1,z_2)A\,. 
\label{eq:fvec}
\end{align}
For $A=i\vp\in \alg{u}(1)$ with $\vp\in \RR$\,, it is explicitly read as 
\begin{align}
(A^*)_{(z_1,z_2)}=(z_1,z_2) i\vp
= i\vp\left(z_1\partial_{z_1}+z_2\partial_{z_2} 
-\ol{z}_1\partial_{\ol{z}_1}-\ol{z}_2\partial_{\ol{z}_2} 
\right)(z_1,z_2)\,. 
\end{align}
Here, the complex conjugate terms are needed for the 
reality condition for $A^*$\,. 

\ms 
\begin{definition}[Fundamental vector field]
The vector field $A^*\in \mathfrak{X}(S^3)$ 
is said to be 
the {\it fundamental vector field} associated with 
$A=i\vp\in \alg{u}(1)$\,. 
It is explicitly defined by 
\begin{align}
(A^*)_{(z_1,z_2)}= i\vp\left(z_1\partial_{z_1}+z_2\partial_{z_2} 
-\ol{z}_1\partial_{\ol{z}_1}-\ol{z}_2\partial_{\ol{z}_2} 
\right)
\qquad \text{at}\qquad (z_1,z_2)\in S^3\,. 
\label{eq:A*}
\end{align}
We denote the map by 
$v: \alg{u}(1)\to  \mathfrak{X}(S^3)$\,, $A\mapsto v(A)=A^{*}$\,. 
\end{definition}

\ms 

For $a\in U(1)$\,, let us denote the right action on $S^3\subset \CC^2$ by 
\begin{align}
R_a : S^3\to S^3\,, \qquad  (z_1, z_2)\mapsto R_a (z_1, z_2)= (z_1, z_2)a \,. 
\end{align}

\ms 

\begin{prop}
\label{prop:om}
The canonical connection $\om$ on $S^3$ in \eqref{eq:cc} satisfies the following conditions: 
\begin{enumerate}
\item[\rm (a)] $\om(A^*)=A$\quad for any \quad$A\in \alg{u}(1)$\,,  
\item[\rm (b)] $(R_a)^*\om=\om$ \quad for any \quad$a\in U(1)$\,. 
\end{enumerate}
\end{prop}

\begin{proof}
(a): By the definitions of $\om$ in \eqref{eq:cc}
and $A^*$ for $A=i\vp$ ($\vp\in \RR$) in \eqref{eq:A*}\,, 
it is directly calculated as 
\begin{align}
\om(A^*)
&=\vev{(z_1, z_2), A^*(z_1, z_2)}
=\vev{(z_1, z_2), i\vp (z_1, z_2)}
\notag \\
&=\left(z_1\ol{z}_1+z_2\ol{z}_2\right) i\vp\,. 
\end{align}
Because of $|z_1|^2+|z_2|^2=1$\,, we obtain $\om(A^*)=i\vp=A$\,.

\ms 
\noindent
(b): By the definition of $\om$ in \eqref{eq:cc}\,, we  see that 
\begin{align}
(R_a)^*\om((z_1,z_2))
&=\om((z_1,z_2)a)=\vev{(z_1,z_2)a, d(z_1,z_2)a}
\notag \\
&=\ol{a} \vev{(z_1,z_2), d(z_1,z_2)} a =\ol{a} \om((z_1,z_2)) a \,. 
\end{align}
Since $\ol{a}a=|a|^2=1$ for $a\in U(1)$\,, we  have $(R_a)^*\om=\om$\,. 
\end{proof}

\ms 

The following proposition means that the connection one-form $\om$ is a projector  
of $TS^3$ onto the subspace, so-called the horizontal subspace. 

\ms 

\begin{defprop}[Horizontal and vertical subspaces]
\label{prop:hor}
Define the subspaces $H_{(z_1,z_2)}$ and $V_{(z_1,z_2)}$ of $T_{(z_1,z_2)}S^3$ by 
\begin{align}
H_{(z_1,z_2)}&={\rm Ker}\, \om=\{~X\in T_{(z_1,z_2)}S^3 ~|~\om(X)=0~\}\,, \notag \\
V_{(z_1,z_2)}&=v(\uu(1)) =\{~(A^*)_{(z_1,z_2)}\in T_{(z_1,z_2)}S^3 ~|~A \in \alg{u}(1)~\}\,.   
\end{align}
These are referred to as the {\rm horizontal} and {\rm vertical subspaces}, respectively. 
Then, it holds that 
\begin{enumerate}
\item[\rm (a)] $T_{(z_1,z_2)}S^3=V_{(z_1,z_2)}\oplus H_{(z_1,z_2)}$ \quad (direct sum)\,, 
\item[\rm (b)] $(R_a)_*H_{(z_1,z_2)}=H_{(z_1,z_2)a}$\quad for any \quad $a\in U(1)$\,. 
\end{enumerate}
\end{defprop}

\ms 
\begin{proof}
(a): By the definition of $v: \alg{u}(1)\to \mathfrak{X}(S^3)$\,, $A\mapsto A^{*}$ in \eqref{eq:A*} and 
the property of $\om$ in \propref{prop:om} (a)\,, the composition map 
\begin{align}
\om \circ v: \uu(1) \to T_{(z_1,z_2)}S^3 \to \uu(1)\,, \quad 
(\om \circ v)(A)=\om((A^*)_{(z_1,z_2)})=A 
\end{align}
is the identity map on $\uu(1)$\,. Hence, the map $v$ is injective, and $\om$ is surjective.  
By the definition of $H_{(z_1,z_2)}={\rm Ker}\, \om$\,, we have the following short exact sequence:
\begin{align}
0\longrightarrow H_{(z_1,z_2)}
\xrightarrow[]{\hspace{2mm} \phantom{\om} \hspace{2mm}}
T_{(z_1,z_2)}S^3 
\xrightarrow[]{\hspace{2mm} \om \hspace{2mm}} \uu(1) \longrightarrow 0 \,. 
\end{align}
By the injectivity of $v : \alg{u}(1)\to T_{(z_1,z_2)}S^3$\,, this sequence splits and we have 
\begin{align}
T_{(z_1,z_2)}S^3 =v(\uu(1))\oplus H_{(z_1,z_2)}=V_{(z_1,z_2)}\oplus H_{(z_1,z_2)}\,. 
\end{align}

\ms 
\noindent
(b): For $a\in U(1)$\,, an arbitrary $Y\in (R_a)_*H_{(z_1,z_2)}$ could be written as $Y=(R_a)_*X=Xa$ with 
$X\in H_{(z_1,z_2)}$\,.  By \propref{prop:om} (b)\,, we have 
\begin{align}
\om(Y)=\om((R_a)_*X)=((R_a)^*\om)(X)=\om(X)=0\,,  
\end{align}
which means that $Y\in H_{(z_1,z_2)a}$ and, hence, $(R_a)_*H_{(z_1,z_2)}\subset H_{(z_1,z_2)a}$\,. 
Conversely, for any $Xa\in H_{(z_1,z_2)a}$ with $X\in T_{(z_1,z_2)}S^3$\,, it holds that 
\begin{align}
0=\om(Xa)=\om((R_a)_*X)=((R_a)^*\om)(X)=\om(X)\,. 
\end{align}
This implies that $X\in H_{(z_1,z_2)}$ and $(R_a)_*H_{(z_1,z_2)}\supset H_{(z_1,z_2)a}$\,. 
This proves that 
\begin{align}
(R_a)_*H_{(z_1,z_2)}= H_{(z_1,z_2)a}\,. 
\end{align}
\end{proof}
\ms 

\begin{remark}
\dpropref{prop:hor} together with \propref{prop:om} are often adopted as the definition 
of the connection one-form 
for the generic principal fiber bundles with the structure group, 
a Lie group $G$\,, where \propref{prop:om} (b) should be replaced by $(R_a)^*\om=a^{-1}\om a$ with $a\in G$\,. 
For details, see \cite{KN, K}\,. 

\end{remark}

\ms 

At the end of this section, we present concrete descriptions of the horizontal and vertical subspaces in terms of local coordinates. 
Set 
\begin{align}
z_1=e^{i\vp_1}\cos\frac{\be}{2}\,, \qquad 
z_2=e^{i\vp_2}\sin\frac{\be}{2}\,, 
\end{align}
where $\vp_1, \vp_2, \be\in \RR$ with 
$0\leq \vp_1, \vp_2< 2\pi$ and $0\leq \be \leq \pi$\,. 
Then, $(z_1,z_2)\in S^3$ because of 
\begin{align}
|z_1|^2+|z_2|^2=1\,. 
\end{align}
The canonical connection $\om$ on $S^3$ is calculated as 
\begin{align}
\om
&=\ol{z}_1dz_1+\ol{z}_2dz_2
=i\cos^2\frac{\be}{2}d\vp_1 
+
i\sin^2\frac{\be}{2}d\vp_2
\notag \\
&=\frac{i}{2}
\left(d\vp+\cos\be d\theta\right)\,. 
\end{align}
In the last equality, we have set 
\begin{align}
\vp=\vp_1+\vp_2\,, \qquad \theta =\vp_1-\vp_2\,. 
\end{align}
The horizontal and vertical subspaces given in
\dpropref{prop:hor} are explicitly described as follows. 

\ms 

\begin{cor}
\label{cor:om}
For the canonical connection $\om$ on $S^3$\,,  
\begin{align}
\om((z_1, z_2))=\frac{1}{2}\left(\ol{z}_1dz_1+\ol{z}_2dz_2
-z_1d\ol{z}_1-z_2d\ol{z}_2\right)
\qquad \text{with}\qquad (z_1,z_2)\in S^3\,, 
\label{eq:cor-om}
\end{align}
the tangent space $T_{(z_1,z_2)}S^3$ is decomposed into 
the direct sum of 
two-dimensional horizontal subspace $H_{(z_1,z_2)}$
and one-dimensional vertical subspace $V_{(z_1,z_2)}$\,. 
Their bases are given by
\begin{align}
H_{(z_1,z_2)}&=\langle~i|z_2|^2(z_1\partial_{z_1}-\ol{z}_1\partial_{\ol{z}_1})
-i|z_1|^2(z_2\partial_{z_2}-\ol{z}_2\partial_{\ol{z}_2})\,, 
\notag \\
&\qquad  -
\frac{|z_2|}{|z_1|}
(z_1\partial_{z_1}+\ol{z}_1\partial_{\ol{z}_1})
+
\frac{|z_1|}{|z_2|}
(z_2\partial_{z_2}+\ol{z}_2\partial_{\ol{z}_2})~\rangle\,,   
\notag \\
V_{(z_1,z_2)}&=\vev{~
i(z_1\partial_{z_1}+z_2\partial_{z_2}
-\ol{z}_1\partial_{\ol{z}_1}
-\ol{z}_2\partial_{\ol{z}_2})
~}\,.  
\label{eq:cor-hv-z12}
\end{align} 
In terms of the spherical coordinates, 
\begin{align}
z_1
=e^{\frac{i}{2}(\vp+\theta)}\cos\frac{\be}{2}\,,
\qquad 
z_2=e^{\frac{i}{2}(\vp-\theta)}\sin\frac{\be}{2}\,, 
\label{eq:cor-ang}
\end{align}
the canonical connection and the horizontal and vertical 
subspaces are simply described as 
\begin{align}
\om((z_1, z_2))
&=\frac{i}{2}\left(d\vp+\cos\be d\theta\right)\,, 
\label{eq:cor-om-ang}
\\
H_{(z_1,z_2)}
&=\vev{~\partial_\theta-\cos\be\partial_\vp\,, 
~\partial_\be~}\,,  
\notag \\
V_{(z_1,z_2)}&=\vev{~\partial_\vp~}\,.  
\label{eq:cor-hv-ang}
\end{align}
\end{cor} 

\ms 

\begin{proof}
The expression of the canonical one-form \eqref{eq:cor-om-ang}
follows from the definition \eqref{eq:cor-om} by 
the coordinate change \eqref{eq:cor-ang}\,. 
We first show \eqref{eq:cor-hv-ang}\,. 
An arbitrary vector $X\in T_{(z_1,z_2)}S^3$ could be written as 
\begin{align}
X=f \partial_\theta+ g \partial_\be+h \partial_\vp
\end{align}
with some real functions $f,g,h$ on $S^3$\,.  
If $X\in H_{(z_1,z_2)}={\rm Ker}\, \om $\,, then 
$\om(X)=0$ implies that $\cos\be f+h=0$\,. 
Hence, any $X\in H_{(z_1,z_2)}$ is written as 
\begin{align}
X=f (\partial_\theta-\cos\be \partial_\vp)+ g \partial_\be\,. 
\end{align}
Since two vectors $\partial_\theta-\cos\be \partial_\vp$
and $\partial_\be$ are linearly independent, they form the basis
of $H_{(z_1,z_2)}$\,. 
On the other hand, the vertical subspace 
$V_{(z_1,z_2)}=v(\alg{u}(1))$ is spanned by the fundamental 
vector field $A^*$ associated with $A=ia\in\alg{u}(1)$ with 
some $a\in \RR$\,. 
By the definition of $A^*$ in \eqref{eq:A*} and the spherical 
coordinates \eqref{eq:cor-ang}\,, it is calculated as  
\begin{align}
A^*= ia \left(z_1\partial_{z_1}+z_2\partial_{z_2} 
-\ol{z}_1\partial_{\ol{z}_1}-\ol{z}_2\partial_{\ol{z}_2} \right)
=2a\partial_\vp\,. 
\end{align}
Hence, $V_{(z_1,z_2)}$ is generated by $\partial_\vp$\,. 

\ms 

Next, the expressions \eqref{eq:cor-hv-z12} are deduced 
from \eqref{eq:cor-hv-ang} via the coordinate change 
\eqref{eq:cor-ang}\,. 
More concretely, substituting the differential operators 
\begin{align}
\partial_\vp&=\frac{i}{2}
\left(z_1\partial_{z_1}+z_2\partial_{z_2} 
-\ol{z}_1\partial_{\ol{z}_1}-\ol{z}_2\partial_{\ol{z}_2} 
\right)\,,
\notag \\
\partial_\theta&=\frac{i}{2}
\left(z_1\partial_{z_1}-z_2\partial_{z_2} 
-\ol{z}_1\partial_{\ol{z}_1}+\ol{z}_2\partial_{\ol{z}_2} 
\right)\,,
\notag \\
\partial_\be&=-\frac{|z_2|}{2|z_1|}\left(
z_1\partial_{z_1}+\ol{z}_1\partial_{\ol{z}_1}\right)
+\frac{|z_1|}{2|z_2|}\left(
z_2\partial_{z_2}+\ol{z}_2\partial_{\ol{z}_2}\right)\,, 
\end{align}
for \eqref{eq:cor-hv-ang}\,, we obtain \eqref{eq:cor-hv-z12}\,.
This completes the proof. 
\end{proof}

\subsection{Horizontal lift and parallel displacement}
\label{sec:para}

Let us now introduce the notion of the {\it horizontal lift}
of a curve in $\cp$ to $S^3$ and 
the {\it parallel displacement} along the horizontal curve. 
The latter is also commonly referred to as {\it parallel transport} 
in the physics and mathematical physics literature.
We shall introduce them for the Hopf fiber bundle endowed with the 
canonical connection. 

\begin{definition}[Horizontal lift]
\label{def:hlift}
Let $c$ be a piecewise differentiable curve of class $C^1$
in $\cp$ described as  
$c=\{~x(t)\in \cp~|~t\in [a,b]~\}\subset \cp \,. $
A {\it horizontal lift} of $c$ is a curve $\wt{c}$ in $S^3$
given by 
\begin{align}
\wt{c}=\{~\wt{x}(t)\in S^3~|~t\in [a,b]~\}\subset S^3\,,   
\end{align}
which satisfies the following two conditions for any $t\in [a,b]$\,;
\begin{itemize}
\item[(i)] $\pi(\wt{x}(t))=x(t)$\,,
\item[(ii)] $\displaystyle{\frac{d}{dt}}\wt{x}(t)
\in H_{\wt{x}(t)}$\,. 
\end{itemize}
If the curve $\wt{c}\subset S^3$ satisfies either (i) or (ii)\,, 
it is said to be a {\it lift} of $c$ or {\it horizontal}, respectively. 
\end{definition}

\ms 

For the Hopf fiber bundle with the 
canonical connection $\om=\vev{(z_1,z_2), d(z_1,z_2)}$\,, 
the condition of horizontal lift of a curve is explicitly described as 
follows. 

\ms 

\begin{prop}
\label{prop:lift}
Let $c$ be a piecewise differentiable curve of class $C^1$ 
in $\cp$ given by 
\begin{align}
c=\{~[z_1(t), z_2(t)]\in \cp~|~t\in [a,b]~\}\subset \cp\,.    
\end{align}
Then,  a curve 
\begin{align}
\wt{c}=\{~(z_1(t), z_2(t))\in S^3~|~t\in [a,b]~\}\subset S^3\,,   
\end{align}
is a horizontal lift of a curve $c$ if it satisfies that 
$\pi((z_1(t), z_2(t)))=[z_1(t), z_2(t)]$ and 
\begin{align}
\ol{z}_1 \dot{z}_1+\ol{z}_2 \dot{z}_2
=0\,.  
\label{eq:ode}
\end{align}
Here, the dot means the derivative with respect to $t$\,, 
i.e.\,, 
$\dot{x}=dx/dt$\,. 
\end{prop}

\begin{remark}
In terms of the parametrization of $S^3$ in \eqref{eq:cor-ang}, 
the horizontal condition can be written as 
\begin{align}
\label{eq:hor-ang}
\ol{z}_1 \dot{z}_1+\ol{z}_2 \dot{z}_2
=\frac{i}{2}
\bigl(\dot{\vp}+\cos\be \dot{\theta}\bigr)=0\,. 
\end{align}
The angular variable $\vp$ in this expression is the Hopf coordinate. 
It should not be
confused with a local fiber coordinate associated with a choice of local
section of the Hopf bundle. 
Precisely, over the local charts $U_1, U_2\subset \cp$ one may write
\begin{align}
\vp+\theta=2\chi_1,\qquad \vp-\theta=2\chi_2, 
\end{align}
where $\chi_1, \chi_2$ are the fiber coordinates relative to the local sections, respectively.
\end{remark}

\ms 
\begin{proof}
To show a curve $\wt{c}\subset S^3$ is a horizontal lift of 
$c\subset \cp$\,, it is sufficient to prove 
the horizontal condition (ii) of \defref{def:hlift}\,, 
\begin{align}
\frac{d}{dt}(z_1(t), z_2(t))\in H_{(z_1(t), z_2(t))}\,, 
\end{align}
as it is a lift of $c$\,, $\pi((z_1(t), z_2(t)))=[z_1(t), z_2(t)]$\,. 
Since we have 
\begin{align}
\om((\dot{z}_1, \dot{z}_2))
= \vev{(z_1, z_2), (\dot{z}_1, \dot{z}_2)}
=\ol{z}_1 \dot{z}_1+\ol{z}_2 \dot{z}_2\,,  
\end{align}
the condition $\om((\dot{z}_1, \dot{z}_2))=0$ is 
equivalent to that 
\begin{align}
\ol{z}_1 \dot{z}_1+\ol{z}_2 \dot{z}_2
=0\,.  
\end{align}
The second equality follows from the coordinate change
\eqref{eq:cor-ang}\,. This completes the proof. 
\end{proof}

\ms 

\begin{defprop}[Parallel displacement]
For a fixed $x_0\in \cp$\,, 
take $y_0\in \pi^{-1}(x_0)\subset S^3$\,. 
Then, there uniquely exists 
a horizontal curve $\wt{c}$ in $S^3$ 
starting from $y_0\in S^3$\,, that is,  
\begin{align}
\wt{c}=\{~\wt{x}(t)\in S^3~|~
\om(\,\dot{\wt{x}}\,)=0\,,~
\wt{x}(a)=y_0\,, ~t\in [a,b]~\}\subset S^3\,.   
\end{align}
The curve $\wt{c}$ is said to be the parallel displacement
of $y_0$ along the curve $c$\,. 
\end{defprop}

\ms 

\begin{proof}
By \propref{prop:lift}\,, the existence and uniqueness of 
a curve $\wt{c}$ reduces to those of a solution of 
the first-order ordinary differential equation of $\vp$ in 
\eqref{eq:ode}\,.  
Since the differential equation \eqref{eq:ode} 
with the initial condition $\wt{x}(a)=y_0$
has a unique solution, the claim follows.
\end{proof}

\ms

\section{Rotation angle as the holonomy}
\label{sec:rot-hol}
\setcounter{equation}{0}

The notion of the Hopf fibration equipped with the canonical connection \eqref{eq:cc}
allows us to describe the geometric phase $\De_g$ in \eqref{eq:mainthm} in terms of 
the holonomy of the Hopf fibration 
(\thmref{thm:hol} in \secref{sec:hol}). 
To state the relation precisely, we need to take into account 
the representation (\defref{def:rho2}), 
\[
\rho_2 : U(1) \longrightarrow U(1),
\qquad
\rho_2(e^{i\varphi}) = e^{2i\varphi}.
\]
The canonical Hopf connection gives the $U(1)$ holonomy,
whereas the geometric phase of the rotating disc is obtained
from this holonomy through the natural 
weight-two representation of the fiber group.
Conceptually, $\rho_2$ may be viewed as a simple
{\it functorial bridge} from the mathematical holonomy of the Hopf fibration 
to the physical rotation angle observed in the model.
Therefore, if $\tau(\ga)\in U(1)$ denotes the
holonomy along the Gauss curve $\ga$, the correct relation is
\[
\rho_2(\tau(\ga))=\exp(i\De_g).
\]
In this sense, the geometric phase $\De_g$ is described 
as the holonomy of the Hopf
fibration, after applying the representation $\rho_2$\,.
Furthermore, we will explain in \secref{sec:cover} that
the information carried by the topological indices 
is retained at the level of the real lift $\RR\to U(1)$\,, 
although it is lost after exponentiation to $U(1)$\,.

\subsection{$U(1)$ holonomy and the geometric phase}
\label{sec:hol}

\subsection*{$U(1)$ holonomy}

Suppose that 
$c=\{~x(t)\in \cp~|~t\in [0,1]~\}\subset \cp$
is a closed curve such as $x(0)=x(1)=x_0$\,.
For some $y_0\in \pi^{-1}(x_0)$\,,  
let $\wt{c}$ be the parallel displacement of $y_0$
along $c$ described by 
\begin{align}
\wt{c}=\{~\wt{x}(t)\in S^3~|~t\in [0,1]~\}\subset S^3\,,  
\end{align}
where $\wt{x}(0)=y_0$ and 
$\pi(y_0)=\pi(\wt{x}(1))=x_0$\,. 
Since both $\wt{x}(0)$ and $\wt{x}(1)$ belong to the same fiber 
$\pi^{-1}(x_0)$\,, there uniquely exists an element 
$\tau(c)\in U(1)$ such that 
\begin{align}
\wt{x}(1)=\wt{x}(0)\cdot \tau(c)\,,  
\end{align}
which is called the {\it holonomy} of $c$
with a base point $x_0$\,.

\ms 

Since $U(1)$ action commutes with the parallel displacement, 
the holonomy is compatible with the concatenation of
based loops:
\begin{align}
\tau(c'\circ c)=\tau(c')\cdot \tau(c)\,, 
\end{align}
where $c'$ is another closed curve based at $x_0$
defined by 
$c'=\{~u(t)\in \cp~|~t\in [0,1]~\}$\,,
and their composition is  
$c'\circ c =\{~v(t)\in \cp~|~t\in [0,1]~\}$
by 
\begin{align}
v(t)=\begin{cases}
x(2t) & (\,0\leq t\leq \frac{1}{2}\,) \\ 
u(2t-1) & (\,\frac{1}{2}\leq t\leq 1\,) \\ 
\end{cases}\,. 
\end{align}
In particular, we have 
$\tau(c^{-1}\circ c)=\tau(c\circ c^{-1})
=1\in U(1)$
where 
$c^{-1}=\{~x(1-t)\in \cp~|~t\in [0,1]~\}$\,. 

\ms

In general, the set $\{\tau(c)\}$ forms a subgroup of 
the structure group $G$\,, which is referred to as 
the {\it holonomy group} with base point $x_0$\,. 
In the case of the Hopf fibration with $U(1)$ structure group, 
the holonomy group is isomorphic to $U(1)$ itself.

\ms
In the subsequent arguments, 
through the diffeomorphism of $S^2$ and $\cp$\,, 
which is  summarized in \appref{app:diffeo}\,, 
we can regard the Gauss curve $\ga$ in \eqref{eq:ga}
as a curve in $\cp$:
\begin{align}
&\ga=\{~[z_1(t)\,,z_2(t)]   \in \cp~|~t\in [0,1]~\}\subset \cp\,,
\\
&\text{where}\quad 
z_1(t)=e^{\frac{i}{2}\theta(t)}\cos\frac{\be(t)}{2}\,, \quad 
z_2(t)=
e^{-\frac{i}{2}\theta(t)}\sin\frac{\be(t)}{2}\,. 
\notag 
\end{align}

\ms 

\begin{prop}\label{prop:hol}
Let $\ga$ be a piecewise smooth closed Gauss curve on
$S^2\simeq\mathbb{CP}^1$
associated with the motion \eqref{eq:motion}, 
and let $\tau(\ga)\in U(1)$ be the
holonomy of the Hopf fibration $S^3\to S^2$ equipped 
with the canonical connection. 
Then, the holonomy is given by 
\begin{align}
\label{eq:prop-hol}
\tau(\ga)
=(-1)^n
\exp\left(
-\frac{i}{2}\int_0^1
\cos\beta(t)\dot\theta(t)\,dt
\right)\,, 
\end{align}
where $n=\theta(1)/2\pi\in\ZZ$  
is the topological number defined in \eqref{eq:top}.
\end{prop}

\begin{remark}
Although the two factors
\begin{align}
(-1)^n
\qquad\text{and}\qquad
\exp\left(
-\frac{i}{2}\int_0^1
\cos\beta(t)\dot\theta(t)\,dt
\right)
\end{align}
on the right-hand side of \eqref{eq:prop-hol}
depend separately on the motion \eqref{eq:motion},
their product depends only on the associated 
Gauss curve $\ga$.
Thus, their combination is an intrinsic quantity 
associated with $\ga$,
as it must be since it gives the holonomy $\tau(\ga)$.
\end{remark}

\begin{proof}[Proof of Proposition \ref{prop:hol}]
Through the diffeomorphism of $S^2$ and $\cp$\,, 
which is  summarized in \appref{app:diffeo}\,, 
we can regard the Gauss curve $\ga$ 
in \eqref{eq:ga}
as a curve in $\cp$:
\begin{align}
\label{eq:ga-cp1}
&\ga=\{~[z_1(t)\,,z_2(t)]   
\in \cp~|~t\in [0,1]~\}\subset \cp\,.
\end{align}

\ms 

We first consider the case where the closed curve $\ga$ 
in \eqref{eq:ga-cp1}
is contained in a single local chart. 
On the chart $U_1$, where $z_1\neq0$ $(\be\neq\pi)$,  
a lift $\wt{\ga}$ can be parametrized as 
\begin{align}
&\wt{\ga}=\{~\si_1(t)e^{i\chi_1(t)}
\in S^3~|~t\in [0,1]~\}\subset S^3
\notag \\
&
\text{with}\quad 
\si_1(t)=\bigl(\cos\frac{\be(t)}{2}\,, \,
e^{-i\theta(t)}\sin\frac{\be(t)}{2}\bigr)\,, 
\end{align}
and it is horizontal if it satisfies the condition
\begin{align}
\om(\tfrac{d}{dt})=(\om_1+id\chi_1)(\tfrac{d}{dt})=0\qquad \text{with}\qquad 
\om_1=\frac{i}{2}(-1+\cos\beta)\,d\theta .
\end{align}
Hence, the holonomy is computed as
\begin{align}
\tau(\ga)
&=e^{i(\chi_1(1)-\chi_1(0))}
=\exp\left(i\int_{\ga} d \chi_1\right) 
=\exp\left(-\int_{\ga} \om_1\right) 
=
\exp\left(\frac{i}{2}\int_{\ga}
(1-\cos\be)\,d\theta \right)
\notag \\
&=\exp\left(\frac{i}{2}\int_0^1 
(1-\cos\beta(t))\,\dot\theta(t)\, dt\right) 
\end{align}
Since the integration of $\theta$ gives the topological number \eqref{eq:top}, 
\begin{align}
\int_0^1 \dot\theta(t)dt =\theta(1)-\theta(0)=2\pi n\,,  
\end{align}
the holonomy reduces to the desired expression, 
\begin{align}
\tau(\ga)=(-1)^n 
\exp\left(
-\frac{i}{2}\int_0^1
\cos\beta(t)\dot\theta(t)\,dt\right)\,. 
\end{align}

\ms 

Similarly, if $\ga$ is included in the chart $U_2$, 
where $z_2\neq0$ $(\be\neq0)$,  
a lift $\wt{\ga}$ can be parametrized as 
\begin{align}
&\wt{\ga}=\{~\si_2(t)e^{i\chi_2(t)}
\in S^3~|~t\in [0,1]~\}\subset S^3
\notag \\
&
\text{with}\quad 
\si_2(t)=\bigl(e^{i\theta(t)}\cos\frac{\be(t)}{2}\,,\, 
\sin\frac{\be(t)}{2}\bigr)\,,
\end{align}
and it is horizontal if it satisfies the condition
\begin{align}
\om(\tfrac{d}{dt})
&=(\om_2+id\chi_2)(\tfrac{d}{dt})=0
\qquad \text{with}\qquad 
\om_2=\frac{i}{2}(1+\cos\beta)\,d\theta .
\end{align}
In  this case, the holonomy is computed as
\begin{align}
\tau(\ga)
&=e^{i(\chi_2(1)-\chi_2(0))}
=\exp\left(i\int_{\ga} d \chi_2\right) 
=\exp\left(-\int_{\ga} \om_2\right) 
=
\exp\left(-\frac{i}{2}\int_{\ga}
(1+\cos\beta)\,d\theta \right)\,. 
\notag \\
&=\exp\left(-\frac{i}{2}\int_0^1 
(1+\cos\beta(t))\,\dot\theta(t) \right) dt 
\end{align}
Evaluating the integration of $\theta$\,, 
we also have the same expression \eqref{eq:prop-hol}. 

\ms

Now consider the general case where $\ga$ is {\it not} 
contained in a single local chart.
Let us decompose the time interval as 
\begin{align}
0=t_0<t_1<\cdots <t_N=1\,, 
\end{align}
such that each associated segment 
\begin{align}
\ga_j=\{~[z_1(t)\,,z_2(t)]\in \cp~|~t\in [t_{j-1}, t_j]~\}
\qquad \text{with}\qquad 
j=1, \cdots, N,  
\end{align}
is contained in one chart $U_{\al_j}$,
where $\al_j\in \{1,2\}$.   
Choose the subdivision so that the initial and final segments 
are contained in the same chart containing the base point
at $t=0$ and $t=1$. 
By combining adjacent segments contained in the same chart,
we may assume that the charts alternate
in the following ordered decomposition of $\ga$, 
\begin{align}
\ga=\ga_1\cup \ga_2\cup \cdots\cup \ga_N\,. 
\end{align}
More precisely, 
we suppose that $N$ is odd and 
that $\ga_j\subset U_1$ for odd $j$, 
while $\ga_j\subset U_2$ for even $j$. 
Thus, both $\ga_1$ and $\ga_N$ are contained in $U_1$.
Note that the subsequent argument
is parallel when they are in $U_2$\,.

\ms

Let us denote the horizontal lift of $\ga_j$ by 
\begin{align}
\wt{\ga}_j
=\{~\si_{\al_j}(t)e^{i\chi_{\al_j}(t)}\in S^3~
|~t\in [t_{j-1}, t_j]~\}
\qquad \text{with}\qquad 
j=1, \cdots, N.   
\end{align}
The holonomy along $\ga$ is then given by 
an ordered product in which the local parallel transport
operators along the segments and 
the transition functions at the switching points 
appear alternately. Since it holds that 
\begin{align}
\si_2(t)=\si_1(t)\psi_{12}(t)\,, 
\qquad 
\psi_{12}(t)=\psi_{21}(t)^{-1}=e^{i\theta(t)}
\end{align}
at the connection points, 
we obtain 
\begin{align}
\label{eq:prop-prof}
\tau(\ga)&=\exp\left(i\int_{\ga_1} d\chi_1 \right)\psi_{21}(t_1)
\exp\left(i\int_{\ga_2} d\chi_2 \right)\psi_{12}(t_2)
\cdots 
\psi_{12}(t_{N-1})
\exp\left(i\int_{\ga_N} d\chi_1 \right)
\notag \\
&=\exp\left(-\int_{\ga_1} \om_1 \right)e^{-i\theta(t_1)}
\exp\left(-\int_{\ga_2} \om_2 \right)e^{i\theta(t_2)}
\cdots 
e^{i\theta(t_{N-1})}
\exp\left(-\int_{\ga_N} \om_1 \right), 
\end{align}
where the second equality follows from 
the horizontal conditions. 
Using the explicit expressions $\om_1$ and 
$\om_2$, we have 
\begin{align}
\tau(\ga)
&=e^{\frac{i}{2}(\theta(t_1)-\theta(t_0))}e^{-i\theta(t_1)}
e^{-\frac{i}{2}(\theta(t_2)-\theta(t_1))}e^{i\theta(t_2)}\cdots 
e^{i\theta(t_{N-1})}e^{\frac{i}{2}(\theta(t_N)-\theta(t_{N-1}))}
\notag \\
&\qquad \times  
\exp\left(-\frac{i}{2}\sum_{j=1}^{N}
\int^{t_j}_{t_{j-1}}\cos\be(t)\, \dot\theta(t)\,dt \right)
\notag \\
&=e^{\frac{i}{2}(\theta(t_N)-\theta(t_0))}
\exp\left(-\frac{i}{2}
\int^{1}_{0}\cos\be(t)\, \dot\theta(t)\,dt \right). 
\end{align}
By the  topological condition \eqref{eq:top}, 
\begin{align}
\theta(t_N)-\theta(t_0)=\theta(1)-\theta(0)
=2\pi n \,. 
\end{align}
Hence, it reduces to 
\begin{align}
\tau(\ga)
&=(-1)^n 
\exp\left(-\frac{i}{2}\int_0^1
\cos\beta(t)\dot\theta(t)\,dt\right)\,. 
\end{align}
Thus, this completes the proof. 
\end{proof}

\begin{remark}
If $\ga$ is simple, then, in terms of the oriented 
spherical areas defined in \eqref{eq:Apm}, 
\propref{prop:hol} 
may equivalently be expressed as
\begin{align}
\tau(\ga)
&=\exp\left(\frac{i}{2}A_+\right)
=\exp\left(-\frac{i}{2}A_-\right).
\end{align}
Thus, the holonomy has a direct geometric interpretation 
as the exponential of one half of the oriented spherical area.
\end{remark}

\subsection*{From $U(1)$ holonomy 
to the geometric phase}

The geometric phase is not identified directly with the 
$U(1)$ holonomy itself, 
but through a representation of the fiber group.
To state the relation precisely, let us define the 
{\it functorial bridge from mathematics} of 
the Hopf fibration {\it to physics} of the rotation angle. 

\begin{definition}
\label{def:rho2}
Define the weight-two representation of the $U(1)$ fiber by 
\begin{align}
\label{eq:wt2}
\rho_2 : U(1) \longrightarrow U(1),
\qquad
\rho_2(u) = u^2.
\end{align}
\end{definition}

\ms 

This weight-two representation allows us to connect 
the geometric phase of our model with the holonomy of 
the Hopf fibration. 

\ms 

\begin{thm}
\label{thm:hol}
Let $\ga$ be a piecewise smooth Gauss curve on 
$\cp$ associated with 
the motion \eqref{eq:motion}\,. The geometric phase 
$\De_g$ and the holonomy $\tau(\ga)$ are related as 
\begin{align}
\rho_2(\tau(\ga))=\exp ({i\De_g})\,. 
\end{align}
\end{thm}

\begin{proof}
By \propref{prop:hol} and 
the geometric phase \eqref{eq:geom},
it holds that 
\begin{align}
\tau(\ga)&=(-1)^n
\exp\left(\frac{i}{2}\De_g\right).
\end{align}
Therefore, applying the weight-two representation 
\eqref{eq:wt2}, we obtain 
\begin{align}
\rho_2(\tau(\ga))
=\tau(\ga)^2
=\exp(i\De_g),
\end{align}
since $(-1)^{2n}=1$.
\end{proof}

\subsection{Geometric phase and a real lift of the 
$U(1)$ holonomy}
\label{sec:cover}

For a simple Gauss curve $\ga$, 
the careful reader might notice that, 
in the description of holonomy, the contributions 
of topological indices $I_\pm\in \{0,1,2\}$ 
for $\De_g=A_+-2\pi I_+$ 
are hidden since 
\begin{align}
\rho_2(\tau(\ga))=\exp(i \De_g)
=e^{i(A_+-2\pi I_+)}=e^{iA_+}\,. 
\end{align}
In this sense, it is useful to consider a real lift of the 
$U(1)$ holonomy in order to retain the real-valued 
information contained in the geometric phase.

\ms 

In other words, $\De_g$ may be regarded as 
an element of the universal covering space
$\RR$ over $S^1\simeq U(1)$\,,
that is  
\begin{align}
&0\longrightarrow 2\pi \ZZ 
\longrightarrow
\RR \overset{p}{\longrightarrow }
U(1) \longrightarrow 1 \,,
\notag \\ 
&\phantom{
0\longrightarrow 2\pi \ZZ \longrightarrow
} 
x\longmapsto \exp(i x)\,. 
\end{align}
This viewpoint is also natural from the Lie-algebraic 
perspective\footnote{
The author thanks Prof.\ Yasukura for pointing this out.}: 
the connection one-form takes values 
in the Lie algebra $\alg{u}(1)=i\RR$, 
and the passage to the $U(1)$ holonomy is made through
exponentiation, which identifies real phases differing by $2\pi$.

\ms 

To state the claim precisely, 
we introduce the lift of 
the representation $\rho_2$ in \defref{def:rho2} 
to the universal covering space of $U(1)$, 
which will be used to relate a lifted Hopf holonomy
to the real-valued geometric phase.

\begin{definition}
\label{def:rhotil2}
Let
\begin{align}
\label{eq:cover}
p:\mathbb R\to U(1),\qquad p(x)=e^{ix}
\end{align}
be the universal covering map. 
Define a natural lift $\wt{\rho}_2$ of 
$\rho_2$ in \defref{def:rho2} by 
\begin{align}
\label{eq:rho-til}
\wt{\rho}_2:\RR\to\RR,
\qquad
\wt{\rho}_2(x)=2x,
\end{align}
satisfying $p\circ \widetilde{\rho}_2=\rho_2\circ p$, which means that   
these maps fit into the commutative diagram,
\[
\begin{CD}
\mathbb R @>{\widetilde\rho_2}>> \mathbb R\\
@V{p}VV @VV{p}V\\
U(1) @>{\rho_2}>> U(1). 
\end{CD}
\]
\end{definition}

\ms 

Let us introduce a {\it lifted holonomy} 
$\wt{\tau}_M(\ga)\in \RR$
satisfying the property 
\begin{align}
\label{eq:p-lift}
p(\wt{\tau}_M(\ga))=e^{i\wt{\tau}_M(\ga)}=\tau(\ga)\,, 
\end{align}
where $p$ is the covering map. 
The subscript $M$ emphasizes that this real lift depends
on the motion \eqref{eq:motion}, 
or equivalently on the choice of regularization,
and not only on the limiting curve $\ga$.
This condition alone, however, determines 
$\wt{\tau}_M(\ga)$ only up to an additive multiple of $2\pi$.  
In the present setting, we choose the following representative,
which is naturally adapted to the expression obtained in 
the proof of \propref{prop:hol}, that is, 
\begin{align}
\label{eq:tautil}
\wt{\tau}_M(\ga)= \pi n-\frac{1}{2}\int_0^1
\cos\beta(t)\dot\theta(t)\,dt\,.
\end{align}
Then, it is easy to see that it matches the condition
\eqref{eq:p-lift}. 
The choice of $\wt{\tau}_M(\ga)$ is understood modulo
$2\pi\ZZ$. In particular, another choice of local trivialization
may give a representative differing from the above 
by an integral multiple of $2\pi$.

\ms 

By the expression of the geometric phase \eqref{eq:geom}, 
it is equal to 
\begin{align}
\wt{\tau}_M(\ga)=\pi n+ \frac{1}{2}\De_g .
\end{align}
Therefore, using the lifted weight-two representation
\eqref{eq:rho-til}, 
we obtain
\begin{align}
\label{eq:lift-tau}
\wt{\rho}_2(\wt{\tau}_M(\ga))
=2\wt{\tau}_M(\ga)=2\pi n+\De_g .
\end{align}
Thus the geometric phase $\De_g$ is not merely 
the image of the Hopf holonomy in $U(1)$, 
but the image of its lifted holonomy under the lifted
representation \(\widetilde\rho_2\). In this lifted formulation, the topological
indices $I_\pm$ are retained:
\begin{align}
\De_g=A_+-2\pi I_+
=-A_-+2\pi I_-.
\end{align}
After exponentiation of \eqref{eq:lift-tau}, 
the topological terms disappear, 
and the result in \thmref{thm:hol} is restored, 
\begin{align}
\rho_2(\tau(\ga))=\exp(i\De_g).
\end{align}

\ms 

%

\begin{remark}
The weight two appearing in $\rho_2$
is not an artifact of the normalization of the angular coordinates.
Its geometric origin is the double covering $SU(2)\to SO(3)$.
The Hopf holonomy takes values in the fiber subgroup 
$U(1)\subset SU(2)$, 
whereas the geometric phase represents the corresponding
physical rotation in $SO(3)$. 
Thus, the weight-two representation $\rho_2(u)=u^2$
reflects precisely this double-covering structure.
%
\end{remark}

\subsection{Geometric phase and the geodesic curvature}
\label{ssec:geom-curv}

Let us denote the geometric phase for 
an arbitrary time $t\in [0,1]$ of the motion \eqref{eq:motion}
by $\De_g(t)$, such that 
$\De_g(0)=0$ and $\De_g(1)=\De_g$\,, namely, 
\begin{align}
\label{eq:De-t}
\De_g(t)= -\int_0^t 
\cos\beta(t)\dot\theta(t)\,dt\,.
\end{align}
This value $\De_g(t)$ 
has a nice geometrical interpretation
in terms of the motion on $S^2$ (\propref{prop:De-t}). 
After preparing some notions of 
local frame and the geodesic curvature
in the subsequent paragraphs,
we shall elucidate this point. 
In the following discussion, the formulas are understood 
on each smooth segment of the Gauss curve away from 
the poles; when necessary, 
they are extended by the regularization introduced in 
\secref{sec:def}.

\subsection*{Local moving frame}

At $\bg(\theta, \be)\in S^2$\,, 
let us introduce the local orthonormal frame by 
\begin{align}
& \ee_1=\frac{\partial \bg}{\partial\theta} \Big/ 
\Big|\frac{\partial \bg}{\partial\theta}\Big|
=\begin{pmatrix} -\sin \theta \\ \cos \theta \\ 0  \end{pmatrix}\,, 
\quad 
\ee_2=-\frac{\partial \bg}{\partial\be} \Big/ 
\Big|\frac{\partial \bg}{\partial\be}\Big|
=\begin{pmatrix} 
-\cos\be \cos \theta \\ 
-\cos\be \sin \theta \\ 
\sin\be  
\end{pmatrix}\,, 
\notag \\
&
\ee_3=\ee_1\times \ee_2 
=\begin{pmatrix}\sin\be \cos\theta \\ \sin\be \sin\theta \\ \cos\be \end{pmatrix}\,,   
\label{eq:e123}
\end{align}
where $\bg(\theta, \be)\in S^2$ is the Gauss vector
\eqref{eq:gauss_a}\,. 
Note that $\ee_3=\bg$\,. 
By the definition, they satisfy 
\begin{align}
&\ee_i \cdot \ee_j=\de_{ij}\,, \qquad 
\ee_i \times \ee_j=\sum_{k=1,2,3} \ep_{ijk}\ee_k
\qquad (i,j=1,2,3)\,, 
\label{eq:ortho}
\end{align}
where $\ep_{ijk}$ is the complete antisymmetric tensor
normalized as $\ep_{123}=1$\,. 
Noting that the vectors $\ee_1$ and $\ee_2$ are always oriented to 
the {\it east} and {\it north}, 
respectively, and $\ee_3$ is to the {\it sky.}  
In particular, $\ee_1$ and $\ee_2$ span the tangent space
$T_{\bg}S^2$\,. 

\ms 

The {\it connection matrix} $(\om_{ij})$ is defined by 
\begin{align}
\begin{pmatrix} d\ee_1 \\ d\ee_2 \\ d\ee_3   \end{pmatrix}
=
\begin{pmatrix}
\om_{11} & \om_{12} & \om_{13} \\ 
\om_{21} & \om_{22} & \om_{23} \\ 
\om_{31} & \om_{32} & \om_{33} 
\end{pmatrix}
\begin{pmatrix} \ee_1 \\ \ee_2 \\ \ee_3   \end{pmatrix}\,. 
\end{align}
By the orthonormal condition \eqref{eq:ortho}\,, it immediately follows that 
$\om$ is anti-symmetric 
\begin{align}
\om_{ij}=-\om_{ji}=d\ee_i\cdot \ee_j
\qquad \text{for}\qquad 
i,j=1,2,3\,. 
\end{align}
It is explicitly calculated as 
\begin{align}
\begin{pmatrix}
\om_{11} & \om_{12} & \om_{13} \\ 
\om_{21} & \om_{22} & \om_{23} \\ 
\om_{31} & \om_{32} & \om_{33} 
\end{pmatrix}
=
\begin{pmatrix}
0 & \cos\be d\theta & -\sin\be d\theta \\ 
-\cos\be d\theta & 0 & d\be \\ 
\sin\be d\theta & -d\be & 0 
\end{pmatrix}\,. 
\label{eq:cmat}
\end{align}
In particular, we see that 
$\om_{12}=\cos\be d\theta$\,.

\ms

\subsection*{Reparameterization from $t$ to $s$}
To describe the motion of disc B, it is convenient to adopt 
the {\it length} of the curve $\ga$ rather than the {\it time} parameter $t\in [0,1]$\,. 
We shall express the length parameter by $s\in [0,L(\ga)]$\,,
where $L(\ga)$ is the length of $\ga\subset S^2$\,, 
and regard $\bg(s)$ as a function of the length $s$ 
rather than the time $t$\,.
The orientation of the curve parametrized by $s$ is induced from 
that of the time parameter $t$ such that    
\begin{align}
\bg(t)\big|_{t=0}=\bg(s)\big|_{s=0}
\qquad \text{and}\qquad 
\bg(t)\big|_{t=1}=\bg(s)\big|_{s=L(\ga)}\,.
\end{align}

\ms 

On an open neighborhood for a fixed $t\in (0,1)$\,, 
where $\bg(t)$ is differentiable and $\bg'(t)\neq0$\,, 
the length parameter $s$ is related to 
the time parameter $t$ via the reparameterization
\begin{align}
s : [0,1] \to [0, L(\ga)]\,, \quad t\mapsto s(t) 
\qquad \text{such that }\qquad 
\left|\bg'(s)\right|= \left|\bg'(t)\right|
\frac{dt}{ds}=1\,.
\label{eq:repara}
\end{align}
By the definition, the speed of $\bg(s)$ is normalized 
as $1$ for the parameter $s$\,. 
We also see that $ds^2$ is the line element of $S^2$. By \eqref{eq:repara} and \eqref{eq:cmat}\,, 
we have 
\begin{align}
ds=\left|\bg'(t)\right| dt\,, 
\qquad 
\bg'(t)=\frac{d\ee_3}{dt}
=\sin\be \frac{d\theta}{dt}\ee_1
-\frac{d\be}{dt}\ee_2\,.  
\end{align}
Therefore, we obtain
\begin{align}
ds^2=\sin^2\be d\theta^2+d\be^2\,. 
\end{align}
This is nothing but the line element of $S^2$ in terms of the local coordinates $(\theta, \be)$\,. 
In the subsequent argument, we reserve $s$ for the length parameter.

\subsection*{Geodesic curvature}

Due to $|\bg'(s)|=1$ and $\bg'(s)\in T_{\bg(s)}S^2
=\text{span}\{\ee_1\,,\ee_2\}$\,, 
we are able to express the tangent vector $\bg'(s)$ by 
introducing an angle 
$\phi$\,,\footnote{In our previous work \cite{MTY}\,, 
the angle $\phi$ was expressed by the symbol $\vp$\,, that is 
$\phi_{\text{[here]}}=\vp_{\cite{MTY}}$\,. 
While $\vp$ is reserved for the $U(1)$ fiber 
coordinate in this article. 
} 
\begin{align}
\bg'(s)=\cos\phi\, \ee_1+\sin\phi\, \ee_2\,, 
\quad \text{where}\quad 
\cos\phi=\sin \be \frac{d\theta}{ds}\,, \quad 
\sin\phi=-\frac{d\be}{ds}\,.  
\label{eq:g-prime}
\end{align}
Figure \ref{fig:frame} illustrates the geometric meaning of 
the angle $\phi$\,.  
Replacing $\phi$ by $\phi+\pi/2$\,, we have the orthogonal vector defined by 
\begin{align}
\bnu(s)=-\sin\phi\, \ee_1+\cos\phi\, \ee_2\,. 
\label{eq:nu}
\end{align}
Note that they satisfy 
\begin{align}
|\bg'(s)|=|\bnu(s)|=1 \,,\qquad  \bg'(s)\cdot \bnu(s)=0\,. 
\label{eq:gvgv-ortho}
\end{align}
The relation between the local frames $\{\ee_1\,,\ee_2\}$
and $\{\bg'(s)\,, \bnu(s)\}$ is presented in 
\figref{fig:frame}. 
\begin{figure}[t]
\centering
\includegraphics[width=10cm]{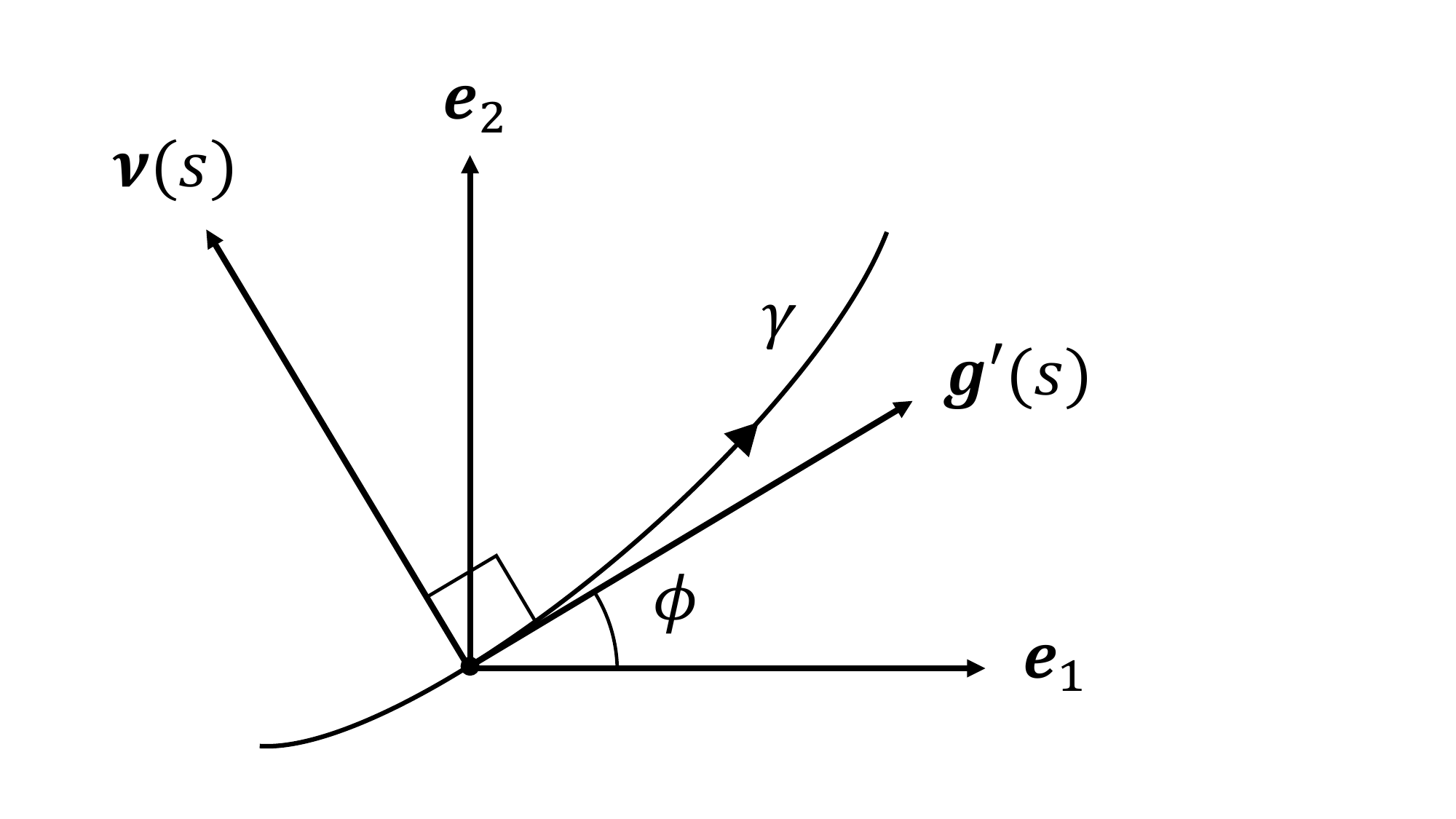}
\caption{The moving frames $\{\bg'(s)\,, \bnu(s)\}$ and 
$\{\ee_1\,,\ee_2\}$ at $\bg(s)\in S^2$\,. 
The vectors $\ee_1$ and $\ee_2$ are oriented to 
east and north, respectively. 
$\phi$ is the angle between the local frames.}
\label{fig:frame}
\end{figure}
Differentiating the second relation, we get 
\begin{align}
\bg''(s)\cdot \bnu(s)+\bg'(s)\cdot \bnu'(s)=0\,. 
\label{eq:gvgv}
\end{align}

\begin{definition}[Geodesic curvature]
\label{def:curv-def}
The {\it geodesic curvature} of $\ga\subset S^2$ at $\bg(s)\in \ga$ is defined by 
\begin{align}
\ka_g(s)= \bg''(s)\cdot \bnu(s)=-\bg'(s)\cdot \bnu'(s)\,.
\label{eq:curv-def}
\end{align}
\end{definition}

By plugging $\bg'(s)$ in \eqref{eq:g-prime} and 
\begin{align}
\bnu'(s)ds 
&= -\cos \phi ( d\phi + \om_{12} )\, \ee_1 
-\sin\phi ( d\phi + \om_{12} )\, \ee_2 
+(-\sin\phi\, \om_{13}+\cos\phi\, \om_{23})\ee_3\,, 
\end{align}
with the geodesic curvature \eqref{eq:curv-def}\,, 
the geodesic curvature is explicitly written as 
\begin{align}
\ka_g(s) ds = d\phi + \om_{12}
=d\phi+\cos\be \,d\theta \,. 
\end{align}
This relation also allows us to express the canonical 
connection of the Hopf fibration directly 
in terms of the geometry of the Gauss curve.

\ms 

\begin{lemma}
\label{lem:om-phi}
The canonical connection $\om$ in \eqref{eq:cor-om-ang}
admits the following expression in terms of the local moving frame along the Gauss curve and its geodesic curvature:
\begin{align}
\om
=\frac{i}{2}\left(d\vp-d\phi+\ka_g(s)ds \right),
\label{eq:om-phi}
\end{align}
where $\vp$ is the Hopf angular 
coordinate in \eqref{eq:cor-ang},
$\phi$ is the angle defined in \eqref{eq:g-prime}, and
$\ka_g$ is the geodesic curvature defined in 
\eqref{eq:curv-def}. 
\end{lemma}

\begin{proof}
The desired expression immediately follows by substituting 
\begin{align}
\cos\be \,d\theta  
=\ka_g(s) ds -d\phi 
\label{eq:con-ka}
\end{align}
for the canonical connection 
$\om$ in \eqref{eq:cor-om-ang}\,. 
\end{proof}

\ms 

On the other hand, the tangent vector 
$d/dt$ of the horizontal lift of the Gauss curve $\ga$
is characterized by the condition $\om(d/dt)=0$\,, 
which is equivalent to \eqref{eq:ode} 
and written by  
\begin{align}
\dot{\vp}-\dot{\phi}+\ka_g(s)\frac{ds}{dt}=0\,. 
\label{eq:horiz}
\end{align}
Integrating this with respect to $t$\,, 
we obtain the following proposition.

\begin{prop}
\label{prop:De-t}
Let $\De_g(t)$ be the geometric phase for 
time $t\in [0,1]$ defined in \eqref{eq:De-t}, 
$\phi(t)$ an angle given 
in \eqref{eq:g-prime}\,,  and 
$\ka_g(s)$ the geodesic curvature of $\ga$ in 
\eqref{eq:curv-def}\,. 
Then we have 
\begin{align}
\De_g(t)&=\phi(t) -\phi(0)
-\int^{s(t)}_{0}\ka_g(s) ds\,. 
\end{align}
\end{prop}

\begin{proof}
By the expression of $\om$ in \eqref{eq:cor-om-ang}
and 
the horizontal condition \eqref{eq:horiz}\,, we have 
\begin{align}
\De_g(t)&= -\int_0^t 
\cos\beta(t)\dot\theta(t)\,dt
=\int_0^t 
\dot\vp(t)\,dt
\notag \\
&
=\int_0^t (\dot{\phi}-\ka_g(s)\frac{ds}{dt})dt 
=\phi(t)-\phi(0)-\int^{s(t)}_{s(0)}\ka_g(s) ds\,. 
\end{align}
As $s(0)=0$ by the definition, 
we obtain the desired relation. 
\end{proof}

\ms 
\begin{cor}
Suppose that the Gauss curve $\ga$ is simple. 
Then, the geometric phase $\De_g$ is expressed as 
\begin{align}
\De_g&=2\pi(1-I_+)
-\int^{L(\ga)}_{0}\ka_g(s) ds 
\label{eq:cor-1}
\\ 
&=2\pi(I_--1)
-\int^{L(\ga)}_{0}\ka_g(s) ds \notag 
\\ 
&=\pi(I_--I_+)
-\int^{L(\ga)}_{0}\ka_g(s) ds \notag 
\,, 
\end{align}
where $I_\pm\in \{0,1,2\}$ are the topological 
indices, namely, the numbers of the corresponding poles
enclosed by the regularized curve $\ga(\ep)$ 
in \eqref{eq:ga-reg}
on its left-hand side.
\end{cor}

\begin{proof}
We only show the first relation \eqref{eq:cor-1}
since the remaining equations follow from $I_++I_-=2$. 
By \propref{prop:De-t} and setting $t=1$\,, 
we get 
\begin{align}
\De_g
&=\phi(1)-\phi(0) 
-\int^{L(\ga)}_{0}\ka_g(s) ds\,, 
\end{align}
since $s(1)=L(\ga)$. 
After applying the same regularization as in the definition of 
$I_\pm$ (and smoothing corners if necessary), 
the total turning of the tangent direction is
\begin{align}
\phi(1) -\phi(0)=2\pi(1-I_+).  
\end{align}
Hence, this completes the proof. 
\end{proof}

\ms 

\begin{remark}
Finally, we comment on the geometric interpretation of
\propref{prop:De-t}.
The angle $\phi$ introduced in \eqref{eq:g-prime}
measures the direction of the tangent vector $\bg'(s)$ 
relative to the local frame $\ee_1$, 
which may be regarded as the needle of a {\it compass}.
Hence, $\phi(1)-\phi(0)$ measures the net rotation of 
disc B relative to the compass.
On the other hand, the integral of the geodesic curvature 
$\ka_g$ measures the total intrinsic turning of the Gauss curve.
Through the horizontal condition $\om=0$ in \eqref{eq:horiz},
these quantities are related to 
the Hopf angular coordinate $\vp$.
Since the geometric phase is related to $\vp$ 
by \eqref{eq:De-t},
\propref{prop:De-t} expresses $\De_g(t)$ 
as a combination of the rotation
relative to the compass and the integrated geodesic curvature.
We expect that this viewpoint will be useful 
in studying the Foucault pendulum \cite{Fou}.\footnote{
Geometric approaches to the Foucault pendulum 
are discussed in
\cite{GR, B, On, Sch, MOWZ, MPM, Mar, MR, BB}.}
\end{remark}

\section*{Acknowledgment}


We thank the Osaka Central Advanced Mathematical Institute,
Osaka Metropolitan University, 
and the Graduate School of Mathematics, Nagoya University, 
for hosting the $32$nd Japan Mathematics Contest and 
the $25$th Japan Junior Mathematics Contest in $2022$, 
at which our model was proposed as one of 
the common problems \cite{JMC}. 
We are grateful to Professors 
Yoshiyuki Koga, 
Mitsutaka Kumakura, 
Yuji Sato, and Yuki Sato 
for helpful discussions on the topics of this article. 
In particular, we thank Professors 
Hiroki Takada and 
Osami Yasukura 
for their collaboration during the early stages of this study. 
This work was supported by 
JSPS KAKENHI Grant Number JP24K06665.

\appendix 
\renewcommand{\theequation}{\Alph{section}.\arabic{equation}}

\section{Diffeomorphism from $S^2$ to $\cp$}
\label{app:diffeo}
\setcounter{equation}{0}

To fix the convention of local coordinates of $S^2$ and $\cp$\,, 
we shall spell out a diffeomorphism of $S^2$ and $\cp$\,.   

\ms 

Two-sphere 
$S^2$ is defined by 
\begin{align}
S^2=\{~(a,b,c)\in \RR^3~|~a^2+b^2+c^2=1~\}
\subset \RR^3\,,
\end{align}
with an atlas $\{(V_1, g_1)\,, (V_2, g_2)\}$\,, where the local 
charts are given by the stereographic projection, namely, 
\begin{align}
&g_1:V_1=\{~(a,b,c)\in S^2~|~c\neq-1~\}\to \CC\,, 
\qquad (a,b,c)\mapsto \frac{a-ib}{1+c}\,, 
\notag \\
&g_2:V_2=\{~(a,b,c)\in S^2~|~c\neq1~\}\to \CC\,, 
\qquad (a,b,c)\mapsto \frac{a+ib}{1-c}\,. 
\end{align}
On $g_1(V_1\cap V_2)=g_2(V_1\cap V_2)=\CC^*$\,, 
the transition map is 
\begin{align}
g_2\circ g_1^{-1}: \CC^*\to \CC^*\,, \quad 
z \mapsto \frac{1}{z}\,. 
\end{align}
In terms of the angular coordinates 
$\be\,, \theta$ $(0\leq \be\leq \pi\,, 0\leq \theta<2\pi)$\,,  
\begin{align}
a=\sin\be\cos\theta\,, \quad 
b=\sin\be\sin\theta\,, \quad 
c=\cos\be\,, 
\end{align}
it can be written as 
\begin{align}
g_1((a,b,c))&=\frac{e^{-i\theta}\sin\be}{1+\cos\be}
=e^{-i\theta}\tan\frac{\be}{2}\,, 
\notag \\
g_2((a,b,c))&=\frac{e^{i\theta}\sin\be}{1-\cos\be}
=e^{i\theta}\cot\frac{\be}{2}\,. 
\end{align}

\ms 

On the other hand, $\cp$ is given by 
\begin{align}
\cp= \{~[z_1, z_2] ~|~(z_1, z_2)\in \CC^2
\backslash \{(0,0)\}~\}\,, 
\end{align}
where the equivalence relation 
$(z_1, z_2)\sim (w_1, w_2)$ is defined by 
\begin{align}
(z_1, z_2)=(\la w_1, \la w_2) \quad \text{with}
\quad \la\in \CC^\times\,.  
\end{align}
An atlas $\{(U_1, f_1), (U_2, f_2)\}$ is defined by 
\begin{align}
f_1 : U_1&=\{~[z_1, z_2] ~|~z_1\neq 0~\}\to \CC\,,  
\quad [z_1, z_2]\mapsto \frac{z_2}{z_1} \,, \notag \\
f_2 : U_2&=\{~[z_1, z_2] ~|~z_2\neq 0~\}\to \CC\,,  
\quad  [z_1, z_2]\mapsto \frac{z_1}{z_2} \,.  
\end{align}
On $f_1(U_1\cap U_2)=f_2(U_1\cap U_2)=\CC^*$\,, 
the transition map reads 
\begin{align}
f_2\circ f_1^{-1} : \CC^*\to \CC^*\,, 
\quad z \mapsto \frac{1}{z}\,. 
\end{align}

\ms 

\begin{prop}
A map $h: S^2 \to \cp$ defined by 
\begin{align}
&h: V_1 \to U_1\subset \cp\,, \quad 
(a,b,c)\mapsto [1:\frac{a-ib}{1+c}]\,, \notag \\
&h: V_2 \to U_2\subset \cp\,, \quad 
(a,b,c)\mapsto [\frac{a+ib}{1-c}:1]\,, 
\label{eq:diffeo}
\end{align}
is a diffeomorphism from $S^2$ to $\cp$\,. 
\end{prop}

\begin{proof}
The map $h$ is smooth since for $i,j=1,2$ it holds that  
\begin{align}
&f_i\circ h\circ g_i^{-1}: \CC\to \CC\,, \quad 
z\mapsto z\,, 
\notag \\
&f_j\circ h\circ g_i^{-1}: \CC^*\to \CC^*\,, \quad 
z\mapsto \frac{1}{z}\,, \quad (i\neq j)\,.  
\end{align}
The inverse map $h^{-1}: \cp\to S^2$ given by 
\begin{align}
&h^{-1}: U_1 \to V_1\subset S^2\,, \quad 
[1:z] \mapsto 
(\,\frac{2{\rm Re}z}{1+|z|^2}\,,
\frac{-2{\rm Im}z}{1+|z|^2}\,,
\frac{1-|z|^2}{1+|z|^2}\,)\,, \notag \\
&h^{-1}: U_2 \to V_2\subset S^2\,, \quad 
[z:1] \mapsto 
(\,\frac{2{\rm Re}z}{1+|z|^2}\,,
\frac{2{\rm Im}z}{1+|z|^2}\,,
\frac{|z|^2-1}{1+|z|^2}\,)
\end{align}
is also smooth. 
Hence, the map $h: S^2 \to \cp$ in \eqref{eq:diffeo}
defines a diffeomorphism from $S^2$ to $\cp$\,. 
This completes the proof. 
\end{proof}



\begin{thebibliography}{99}



\bibitem{Pan}
S.~Pancharatnam,
``Generalized theory of interference, and its applications,''
Proc. Indian Acad. Sci. A \textbf{44} (1956) no.5, 247-262. 

\bibitem{Lon}
H.~C.~Longuet-Higgins,  U.~Opik, M.~H.~L.~Pryce, and R.~A.~Sack,
"Studies of the Jahn-Teller Effect. II. The Dynamical Problem,"
Proceedings of the Royal Society of London Series A,
1958, feb, vol. 244, No. 1236, pp. 1-16. 

\bibitem{han}
J.~H.~Hannay, 
``Angle variable holonomy in adiabatic excursion of 
an integrable Hamiltonian,''
1985 J. Phys. A: Math. Gen. 18 221. 



\bibitem{Berry1}
M.~V.~Berry,
``Quantal phase factors accompanying adiabatic changes,''
Proc. Roy. Soc. Lond. A \textbf{392} (1984), 45-57. 

\bibitem{Berry2}
M.~V.~Berry,
``The Quantum Phase, Five Years After,''
original contribution to \cite{SW}\,. 


\bibitem{SW}
A.~Shapere and F.~Wilczek, 
"Geometric Phases in Physics,"
Advanced Series in Mathematical Physics Vol.5, 
World Scientific, Singapore, 1989.


\bibitem{JMC}
T.~Matsumoto, H.~Takada, and O.~Yasukura,
The common problem: ``Rotation angles of the rotating disc'',
The 32nd Japan Mathematics Contest and the 25th Japan Junior Mathematics Contest, 2022,
organized by the Osaka Central Advanced Mathematical Institute,
Osaka Metropolitan University, and the Graduate School of Mathematics,
Nagoya University.
Available at: \url{https://sites.google.com/view/jmathcon/past_problems}


\bibitem{MTY}
T.~Matsumoto, H.~Takada and O.~Yasukura,
``Rotation Angles of a Rotating Disc --A Toy Model Exhibiting the Geometric Phase--,'' 
FORMA, 2026, Volume 41, Issue 1, Pages 13-30, 
Released on J-STAGE May 14, 2026, Online ISSN 2189-1311, 
Print ISSN 0911-6036, 
\url{https://doi.org/10.55653/forma.2026.001.002}, 
\url{https://www.jstage.jst.go.jp/article/forma/41/1/41_13/_article/-char/en},
[arXiv:2505.16749 [math-ph]].




\bibitem{KN}
S.~Kobayashi and K.~Nomizu, 
``Foundations of Differential Geometry Volume I,II'' 
(1963, 1969), John Wiley \& Sons. 

\bibitem{K}
S.~Kobayashi, 
``Differential geometry of connections and gauge theory,''
Shokabo Tokyo, May 15, 1989, 9th edition (in Japanese). 
%


\bibitem{EGH}
T.~Eguchi, P.~B.~Gilkey and A.~J.~Hanson,
{\it ``Gravitation, Gauge Theories and Differential Geometry,''} 
Phys. Rept. \textbf{66} (1980), 213. 


\bibitem{hopf}
H.~Hopf, 
``\"Uber die Abbildungen der dreidimensionalen Sph\"are 
auf die Kugelfl\"ache,'' Math. Ann. 104, 637–665 (1931). 















%










\bibitem{Fou}
Jean Bernard L\'eon Foucault, 
``D\'emonstration physique du mouvement de rotation de la 
Terre au moyen du pendule,'' 
Comptes rendus hebdomadaires des s\'eances de 
l'Acad\'emie 
des sciences, Tome 32, 1851 (p.135-138).

\bibitem{GR}
L.~E.~Goodman, A.~R.~Robinson,  
"Effect of Finite Rotations on Gyroscopic Sensing Devices." 
ASME. J. Appl. Mech. June 1958; 25(2): 210–213. 


\bibitem{B}
K.~Blankinship, 
"A new kinematic theorem for rotational motion," 
PLANS 2004. Position Location and Navigation Symposium
 (IEEE Cat. No.04CH37556), Monterey, CA, USA, 2004, 
 pp. 285-295.


\bibitem{On}
H.~K.~Onnes,  
``Nieuwe Bewijzen voor de aswenteling der aarde.'' 
(Ph.D. dissertation)
Groningen: Wolters, 1879. 

\bibitem{Sch}
E.~O.~Schulz-DuBois,
"Foucault Pendulum Experiment by Kamerlingh Onnes and degenerate perturbation theory," 
(1970), Am. J. Phys. 38 (2): 173. 


\bibitem{MOWZ}
J.~E.~Marsden, 
O.~M.~O'Reilly, 
F.~J.~Wicklin, 
B.~W.~Zombros, 
``Symmetry, Stability, Geometric Phases, and Mechanical Integrators (Part I, II)''
(1991), Nonlinear Science Today 1 (1) 4–11 and (2) 13-21.

\bibitem{MPM}
F.~Monroy-Perez, Anzaldo-Meneses, 
``Study of the Foucault pendulum within the geometric control theory perspective,'' 
Cybernetics and Physics, Vol. 1, No. 2, 2012, pp. 89-95.

\bibitem{Mar}
J.~E.~Marsden,  
``Lectures on Mechanics,'' 
London Mathematical Society Lecture note series, 174, 
Cambridge University Press (1992).

\bibitem{MR}
J.~E.~Marsden and T.~S.~Ratiu, 
``Introduction to Mechanics and Symmetry,'' 
2nd ed., Texts in Applied Mathematics 17, 
Springer-Verlag, New York, 1999.


\bibitem{BB}
Jens von Bergmann, HsingChi von Bergmann, 
``Foucault pendulum through basic geometry,'' 
Am. J. Phys. 1 October 2007; 75 (10): 888–892. 










\end{thebibliography}
\end{document}